\documentclass[a4paper,10pt,twoside]{article}

\usepackage[latin1]{inputenc} 
\usepackage[T1]{fontenc} 
\newlength{\minipagewidth}
\usepackage{a4wide}
\usepackage{amsmath}
\usepackage{amssymb}
\usepackage{a4wide}
\usepackage{graphicx} 
\usepackage{color}
\usepackage{epic}

\newcommand{\hA}{{\bf 1}_{\rm A}}
\newcommand{\hB}{{\bf 1}_{\rm B}}
\newcommand{\AB}{{AB}}
\newcommand{\cE}{{\cal E}}
\newcommand{\dps}{\displaystyle}

\newtheorem{algo}{Algorithm}[section]

\title{Path Sampling with Stochastic Dynamics: Some new Algorithms}
\author{Gabriel Stoltz$^{1,2}$\\
\footnotesize{1: CERMICS, Ecole Nationale des
  Ponts et Chauss\'ees (ParisTech),}\\ \footnotesize{6 \& 8 Av. Pascal,
  77455 Champs-sur-Marne, France.} \\
\footnotesize{2: CEA/DAM Ile-de-France,
BP 12, 91680 Bruy\`eres-le-Ch\^atel, France.}\\
\footnotesize{stoltz@cermics.enpc.fr} 
}

\begin{document}

\maketitle

\begin{abstract}
We propose here some new sampling algorithms for Path Sampling
in the case when stochastic dynamics are used. In particular, we present
a new proposal function for equilibrium sampling of paths with a Monte-Carlo dynamics (the so-called ``brownian tube'' proposal). This proposal is based on the continuity of the dynamics with
respect to the random forcing, and generalizes all previous
approaches when stochastic dynamics are used. The efficiency of this proposal is demonstrated using some
measure of decorrelation in path space. 
We also discuss a switching strategy that allows to
transform ensemble of paths at a finite rate while remaining at
equilibrium, in contrast with the usual Jarzynski like switching. This switching is very interesting to  
sample constrained paths starting from unconstrained paths, or to
perform simulated annealing in a rigorous way. 
\end{abstract}

\maketitle

\section{Introduction}

The behavior of many systems in the fields of physics, chemistry and biology, is
dictated by rare but important transitions between metastable
states. Usually, only some local exploration of the metastable sets can be
performed, and it is very difficult to study the transitions by
resorting to straightforward simulations - using for
example molecular dynamics or kinetic Monte-Carlo. The Transition Path
Sampling (TPS) formalism, first proposed in~\cite{Pratt86} and further
developped in~\cite{DBCC98} (see also~\cite{BDGC00,DBG02} for extensive
reviews), is a strategy to sample only those paths that lead to a
transition. It also gives some information on the transition kinetics,
such as the rate constant as a function of time or the activation
energies~\cite{DB04}. Recent practical and theoretical developments (such as Transition
Interface Sampling~\cite{vEMB03,vEB05}) are still aiming at increasing the
power of the method. State of the art applications of path sampling, such
as~\cite{JB06}, now involve as much as $15,000$ atoms with paths as long
as $10$~ns. 

Recently, relying on the Jarzynski formula~\cite{jarzPRL,jarz} (roughly
speaking, an exponential average over the works performed during the
switching from an initial to a final state), path sampling techniques
have also been used to compute free energy differences more
efficiently~\cite{Sun03,ZY,ODG05} by precisely enhancing the paths that
have the larger weights (which correspond to the unlikely lower work
values).

Many path sampling studies (especially TPS studies) have used
deterministic dynamics (Path sampling in the NVE ensemble has already
been thoroughly studied, see~\cite{DBG02} for a review). However, path
sampling with stochastic dynamics is of great interest for
nonequilibrium simulations~\cite{CC01}. Besides, some models are
stochastic by nature (see {\it e.g}~\cite{AFR06} where the authors
consider a model system of protein pulling in implicit solvent, and a
chemical reaction simulated with kinetic Monte Carlo). Finally, we
believe that there is room for improvement in the path sampling
techniques for stochastic dynamics. We therefore restrict ourselves to
the stochastic setting in this study.

To this date, the usual equilibrium sampling of paths with stochastic
dynamics is done either with the usual shooting dynamics inspired from
the corresponding algorithm for deterministic paths~\cite{DBG02}; or
with the so-called "noise history" algorithm introduced in~\cite{CC01},
which relies on the description of paths as a starting point and the
sequence of random numbers used to generate the trajectory. It is one of
our aims here to relate both strategies and generalize them by
introducing a new way to propose paths: namely by generating random
numbers correlated with the ones used to generate the previous
path. When the correlation is zero, the usual shooting dynamics is
recovered. When the correlation is one everywhere except for some index
along the path where it is zero, the noise-history algorithm is
recovered. This generalization may be useful for example when the
dynamics are too diffusive (Langevin dynamics in the high friction
limit) since the shooting dynamics are inefficient in this limit; or to enhance the decorrelation of the paths generated using the noise history algorithm. 

We also consider nonequilibrium sampling of paths, using some switching
dynamics on paths~\cite{GD04}, inspired from the now well-known
Jarzynski out-of-equilibrium switching in phase-space~\cite{jarzPRL,jarz}. This switching
can be performed whatever the underlying dynamics on paths. It 
can be used to transform a sample of unconstrained paths to
reactive paths (ending up in some given region). This approach was
already followed in~\cite{GD04}, and allows to compute rate
constants. However, the final sample of paths is very degenerate, and
cannot be used as a reliable equilibrium sample of reactive paths. In
the same vein, one could imagine doing simulated annealing on paths
(simulated tempering on paths has already been investigated
in~\cite{VS01}), in order to obtain typical transition paths at
temperatures where direct sampling is not feasible. However, unless the
annealing process is very slow, the final sample is usually not
correctly distributed. We therefore also present the application to path
sampling of a birth/death process, the so-called "Interacting Particle
System" (IPS), already used in~\cite{RS06} in the field of molecular
dynamics to compute regular phase space properties. This methodology is widely used in the fields of Quantum Monte
Carlo~\cite{caffarel, Rousset06} or Bayesian Statistics, where it is
referred to as Sequential Monte Carlo~\cite{arnaudbook}. It allows,
through some selection of the paths during the nonequilibrium switching
at a finite rate, to precisely reequilibrate the paths distribution at
all times. Such a reequilibration is of paramount importance for the end
sample to be distributed according to the canonical measure on
paths. Besides, since the sample of paths follows the canonical
distribution at all times, the properties of interest can be computed in
a single simulation for a whole range of values. For example, the rate
constant could be obtained for a whole range of temperatures, which
allows to compute the activation energy following the method presented in~\cite{DB04}.

The paper is organized as follows. We first present the path ensemble in
section~\ref{sec:path_ensemble}, and turn to equilibrium sampling of
paths in section~\ref{sec:eq}. We introduce in particular in
section~\ref{sec:tube} the "brownian tube" proposal function which
generalizes the previous algorithms for path sampling with stochastic
dynamics, and compare this new proposal functions to the previous ones
using some two-level sampling indicators (for the local sampling, see
section~\ref{sec:local_sampling} where an abstract measure of diffusion
in path space is introduced). Finally, we present in
section~\ref{sec:noneq} the switching dynamics on paths, with the IPS
extension enabling a reequilibration of the paths distribution at all
times, even when the switching is done at a finite rate (see section~\ref{sec:IPS}).

\section{The path ensemble with stochastic dynamics}
\label{sec:path_ensemble}

\subsection{The canonical measure on discretized paths}

We consider a system of $N$ particles, with mass matrix $M={\rm Diag}(m_1,\dots,m_N)$, described by a configuration variable $q=(q_1,\dots,q_N)$, and a momentum variable $p=(p_1,\dots,p_N)$. The dimension of the space is denoted by $d$, so that $q_i,p_i \in \mathbb{R}^d$ for all $1 \leq i \leq N$.
We consider stochastic dynamics of the form
\begin{equation}
\label{SDE}
dX_t = b(X_t) \, dt + \Sigma dW_t,
\end{equation}
where the variable $X_t$ represents either the configurational part $q_t$, or the full phase space variables $(q_t,p_t)$. The function $b$ is the force field, the matrix $\Sigma$ is the magnitude of the random forcing, and $W_t$ is a standard Brownian motion (the dimension of $W_t$ depending on the dynamics used).

We restrict ourselves in this study to the most famous stochastic dynamics used in practice, namely the Langevin dynamics
\begin{equation}
\label{e:Pqdyn}
\left \{ \begin{array}{cl}
d q_{t} & = \displaystyle M^{-1} \, p_t \, dt, \\
d p_{t} & = \displaystyle -\nabla V(q_t) \, dt - \gamma
M^{-1} p_{t} \, dt + \sigma \, d W_t, \\
\end{array} \right.
\end{equation}
where $W_t$ denotes a standard $dN$-dimensional Brownian motion, and
with the fluctuation-dissipation relation $\sigma^2 = 2
\gamma/\beta$. In this case, the variable $x=(q,p)$ describes the system and the energy is given by the Hamiltonian $E(x) = H(q,p) =
V(q) + \frac{1}{2} p^T M^{-1} p$. Some studies (see {\it
  e.g.}~\cite{ZY}) however resort to the overdamped Langevin dynamics
\[
dq_t = -\nabla V(q_t) \, dt + \sqrt{\frac2\beta} \, dW_t,
\]
in which case $x = q$ and $E(x) = V(q)$.
The ideas presented in the sequel can of course be straightforwardly extended to this case.

In practice, the dynamics have to be discretized. Considering a time step $\Delta t$ and a trajectory length $T = L \Delta t$, a discrete trajectory is then defined through the sequence
\[
x = (x_0,\dots,x_L).
\]
Its weight is
\begin{equation}
\label{canonical_path_discretized_general}
\pi(x) = Z_L^{-1} \rho(x_0) \prod_{i=0}^{L-1} {\rm p}(x_i,x_{i+1}),
\end{equation}
where $\rho(x_0) = Z^{-1}_0 {\rm e}^{-\beta E(x_0)}$ is the Boltzmann
weight of the initial configuration, ${\rm p}(x_i,x_{i+1})$ is the
probability that the system is in the state $x_{i+1}$ conditionally that
it starts from $x_i$, and $Z_L$ is a normalization constant. This
conditional probability depends on the discretization of the dynamics
used.  

Denoting by $\hA(x), \hB(x)$ the indicator functions of some sets $A,B$
defining respectively the initial and the final states, the probability
of a given {\it reactive} path between the sets $A$ and $B$ is then
\begin{equation}
\label{canonical_path_discretized}
\pi_{AB}(x) = Z_\AB^{-1} \hA(x_0) \rho(x_0) \prod_{i=0}^{L-1} {\rm p}(x_i,x_{i+1}) \hB(x_L).
\end{equation}
Transition Path Sampling~\cite{DBCC98,DBG02} aims at sampling the
measure\footnote{Notice that the measure $\pi_{AB} \equiv
  \pi_{AB}^{L,\Delta t}$ depends in fact explicitely on the length of
  the paths, and of the time steps used in practice.} $\pi_{AB}$, using
in particular Monte-Carlo moves of Metropolis-Hastings type. 

\subsection{Discretization of the dynamics}

We present here a possible discretization of the Langevin dynamics, and the corresponding transition probability ${\rm p}(x_i,x_{i+1})$.
This discretization, called ``Langevin Impulse''~\cite{skeel}, relies on an operator splitting
technique, and is more appealing from a theoretical viewpoint than
previous discretizations (such as the BBK algorithm~\cite{BBK}, or
schemes proposed in~\cite{AT}). For particles of equal masses (up to a rescaling of time, $M = {\rm Id}$; the extension to the general case is straightforward), the numerical scheme we use here reads~\cite{skeel}: 
\begin{equation}
\label{splitting}
\left \{ \begin{array}{c@{ \ = \ }l}
\dps p_{i+1/2} &  \dps p_i - \frac{\Delta t}{2} \nabla V(q_i), \\
\dps q_{i+1} & \dps q_i + c_1 \, p_{i+1/2} + U_{1,i}, \\
\dps p_{i+1} & \dps c_0 \, p_{i+1/2} - \frac{\Delta t}{2} \nabla V(q_{i+1}) + U_{2,i}, \\
\end{array} \right.
\end{equation}
with 
\[
c_0 = \exp(-\gamma \Delta t), \quad c_1 = \frac{1-\exp(-\gamma \Delta t)}{\gamma}.
\]
The centered gaussian random variables $(U_{1,i},U_{2,i})$ with $U_{k,i}=(u_{k,i}^1,\dots,u_{k,i}^{dN})$ are such that
\[
\mathbb{E}\left [ (u_{1,i}^l)^2 \right ] = \sigma_1^2, \quad
\mathbb{E}\left [ (u_{2,i}^l)^2 \right ] = \sigma_2^2, 
\quad \mathbb{E}\left [ u_{1,i}^l \cdot u_{2,i}^l \right ] = c_{12} \sigma_1 \sigma_2,
\]
with
\[
\sigma_1^2 = \frac{\Delta t}{\beta \gamma} \left ( 2 - \frac{3-4{\rm
      e}^{-\gamma \Delta t} + {\rm e}^{-2\gamma \Delta t}}{\gamma
    \Delta t} \right ), \quad \sigma_2^2 = \frac{1}{\beta} 
\left ( 1 - {\rm e}^{-2\gamma \Delta t} \right ), \quad c_{12} \sigma_1
\sigma_2 = \frac{1}{\beta \gamma} \left ( 1 - {\rm e}^{-\gamma \Delta t}
\right )^2.
\]
In practice, the random vectors $(U_{1,i},U_{2,i})$ are computed from standard gaussian random numbers $(G_{1,i},G_{2,i})$ with $G_{k,i}=(g_{k,i}^1,\dots,g_{k,i}^{dN})$:
\begin{equation}
u_{1,i}^l = \sigma_1 \, g_{1,i}^l, \quad u_{2,i}^l = \sigma_2 \left ( \sqrt{1-c_{12}^2} \, g_{2,i}^l + c_{12} \, g_{1,i}^l \right).
\end{equation}
We will always denote by $G$ standard gaussian random vectors in the sequel, whereas the notation $U$ refers to non-standard gaussian random vectors.

Denoting by
\[
{\rm d}_1 \equiv {\rm d}_1((q_{i+1},p_{i+1}),(q_i,p_i)) = \left|q_{i+1}-q_i-c_1 \, p_i + c_1 \frac{\Delta t}{2} \nabla V(q_i)\right|, 
\]
\[
{\rm d}_2 = {\rm d}_2((q_{i+1},p_{i+1}),(q_i,p_i)) = \left|p_{i+1}-c_0 \, p_i + \frac{\Delta t}{2} \left( c_0 \nabla V(q_i) + V(q_{i+1}) \right) \right|,
\]
the conditional probability ${\rm p}((q_{i+1},p_{i+1}),(q_i,p_i))$ to be in the state $x_{i+1}=(q_{i+1},p_{i+1})$ starting from $x_i=(q_i,p_i)$ reads
\begin{equation}
\label{splitting_conditional}
{\rm p}(x_{i+1},x_i) = Z^{-1} \exp \left [ - \frac{1}{2 (1 - c_{12}^2)} \left ( \left (\frac{{\rm d}_1}{\sigma_1}\right)^2 + \left (\frac{{\rm d}_2}{\sigma_2}\right)^2 - 2c_{12}\left (\frac{{\rm d}_1}{\sigma_1}\right)\left (\frac{{\rm d}_2}{\sigma_2}\right) \right ) \right ]
\end{equation}
where the normalization constant is $Z = \left ( 2 \pi \sigma_1 \sigma_2 \sqrt{1-c_{12}^2} \right )^{-d}$.

\section{Equilibrium sampling of the path ensemble}
\label{sec:eq}

The most popular way to sample paths is to resort to a
Metropolis-Hastings scheme~\cite{MRRTT53,Hastings70}. 
Other approaches may be considered in some cases , see~\cite{DBG02} for a
review of alternative approaches. Those approaches however require
some force evaluation (see {\it e.g.}~\cite{DBCC98} for a Langevin dynamics in phase space in the case of a toy two-dimensional problem). But the force exerted on a path is proportional to $\nabla (\ln \pi)$, and is  
difficult to compute in general since it requires the evaluation of second derivatives of the potential 
in conventional phase space.

We first recall the general structure of the Metropolis-Hastings
algorithm, and precise some of its specifities, especially when sampling
reactive paths. We then recall a usual technique to propose paths in
section~\ref{sec:shooting}, and generalize it in
section~\ref{sec:tube}. We finally propose some benchmarks to compare
the efficiencies of all these proposal functions. 

\subsection{Metropolis-Hastings sampling techniques for path sampling}
\label{sampling_algo}

For a probability measure $\pi$ on the discretized path ensemble (such as~(\ref{canonical_path_discretized_general}) or~(\ref{canonical_path_discretized})), a Metropolis-Hastings scheme is defined as a Markov chain with transition probability kernel
\begin{equation}
\label{MH_path}
P(x,dy) = r(x,y) {\cal P}(x,y) \ dy + \left ( 1 - \int r(x,y') {\cal P}(x,y') \ dy' \right ) \delta_x, 
\end{equation}
where the density $r(x,\cdot)$ is given by
\begin{equation}
\label{MH_AR}
r(x,y) = \min \left ( 1, \frac{\pi(y) {\cal P}(y,x)}{\pi(x) {\cal P}(x,y)} 
\right ).
\end{equation}
The function ${\cal P}$ is the proposal function (It is more commonly called 'generation probability' in the field of molecular simulation).
In words, the path $y$ is proposed with probability ${\cal P}(x,y)$ from $x$, and accepted with probability $r(x,y)$, rejected otherwise.

The measure $\pi$ is by construction an invariant measure of the
corresponding Markov chain, and $P$ is the transition kernel. For all $n
\geq 0$, $P^n(x,{\cal A})$ is the probability to reach the set ${\cal
  A}$ in $n$ steps starting from $x$ (recall that $P^n(x,\cdot)$ is a
probability measure for all $n \geq 0$).
If for all $x,y$, there exists $n \geq 1$ such that $P^n(x,y) > 0$, and the chain is aperiodic,
then the Markov chain is ergodic~\cite{MT} (We refer to~\cite{CLS05} for
an introduction to ergodicity issues for sampling schemes in the field
of Molecular Dynamics).

The key point in all Metropolis-Hastings schemes is to find an efficient
proposal function. In particular, there is always a trade-off
between the acceptance and the decorrelation rate of the Markov
chain. Indeed, if the acceptance rate is low, the obtained sample is
degenerate, and not statistically confident. On the other hand, to
increase the acceptance rate, more correlated iterations can be used. In
this case the method is more likely to remain trapped in local minima,
and the numerical ergodicity rate may be slow.
In many situations, the optimal acceptance rate is around $1/2$. This heuristic rule can be made rigorous in some situations (see e.g.~\cite{RR97} where the optimal acceptance rate is shown to be $0.574$ for a specific Metropolis-Hastings scheme based on a Euler-Maruyama proposition, in the limit when the dimension of the phase-space goes to infinity).

In the case of reactive paths, a study of the acceptance rate asks to
decompose the acceptance/rejection procedure in two successive steps:
(i) the proposition of a path starting from $A$ and going to $B$;
(ii) the acceptance or rejection of such a path according to the Metropolis-Hastings scheme.
The difficult step is the first one, since paths bridging $A$ and $B$
are only a (small) 
subset of the whole path space. In particular, diffusive dynamics such as the overdamped Langevin
dynamics are often not convenient to propose bridging paths; the situtation is however better for dynamics with some inertia, such as the Langevin dynamics. When the paths are constructed using deterministic dynamics (NVE case), some studies have shown that the optimal acceptance rate is about 40 \% for the cases under consideration~\cite{DBG02}.

For path sampling with stochastic dynamics, the "shooting"
proposal function is classically used~\cite{DBG02}.
However, even for moderate values of the friction coefficient $\gamma$ in the
Langevin dynamics, this proposal function may have low acceptance rates, 
especially if the dimension of the system is high or/and the barriers
to cross are large. An alternative way of proposing paths, relying on
the so-called ``noise history'' of the paths~\cite{CC01} ({\it i.e.} the
sequence of random numbers used to generate the trajectory from a given
starting point) is to change only one of the random numbers used and to
keep the others. In this case, a high acceptance rate is expected, but
the paths generated may be very correlated. 

A natural generalization of
both approaches is to rely on the continuity of the dynamics with
respect to the random noise forcing, and to propose a new trajectory by
generating new random numbers correlated with the previous one. We call
this approach the ``brownian tube'' proposal. In this
case, an arbitrary acceptance rate can be reached, and there is room for
optimizing the parameters in order to really tune the efficiency of the sampling. 

\subsection{The shooting proposal function}
\label{sec:shooting}

The shooting technique described in~\cite[section~3.1.5]{DBG02} consists in the three following steps, starting from a path $x^n$:
\begin{itemize}
\item select an index $0 \leq i \leq L$ according to discrete probabilities $(w_i)_{0 \leq i \leq L}$ (for example a uniform probability distribution can be considered, unless one wants to increase trial moves starting from certain regions, for example the assumed transition region);
\item generate a new path $(y_{i+1},\dots,y_L)$ forward in time, using
  the stochastic dynamics~(\ref{e:Pqdyn}), with a new set of
  independently and identically
  distributed (i.i.d.) gaussian random numbers $(U_j^{n+1})_{i+1 \leq j \leq L-1}$;
\item generate a new path $(y_{i-1},\dots,y_0)$ backward in time, using a discretized "backward" stochastic dynamics correponding to~(\ref{e:Pqdyn}), with a new set of i.i.d. gaussian random numbers $(\overline{U}_j^{n+1})_{0 \leq j \leq i-1}$;
\item set $x^{n+1} = y$ with probability $r(x^n,y)$, otherwise set $x^{n+1} = x^n$. 
\end{itemize}

The ``backward'' part of the trajectory can be computed using some backward integration (resorting to negative time steps), but the associated schemes are often unstable~\cite{stoltz_PhD}. Therefore, a more appropriate method is to resort to {\it time
  reversal}: The forward dynamics are used to generate the points $y_i$
from $y_{i+1}$ in a time-reversed manner. This means that variables odd
with respect to time reversal (such as momenta) are inverted, and
variables even with respect to time reversal (such as positions) are
kept constant. Denoting by ${\cal S}$ the reversal operator, ${\cal S}y_i=(q_i,-p_i)$ when $y_i=(q_i,p_i)$ for Langevin dynamics. The usual one-step integrator $\Phi_{\Delta t}$ is then considered to integrate the corresponding trajectory: 
\[
y_i = ({\cal S} \circ \Phi_{\Delta t} \circ {\cal S}) y_{i+1}.
\]
The time-reversed conditional probability ${\rm \bar{p}}_{\rm TR}(y_{i+1},y_i)$ to go from $y_{i+1}$ to $y_i$ is
\[
{\rm {\bar p}}(y_{i+1},y_i) = {\rm \bar{p}}_{\rm TR}(y_{i+1},y_i) = {\rm p}({\cal S}y_{i+1}, {\cal S}y_i).
\]
We will always denote in the
sequel the random numbers used in this process by $\bar{U}$.
The probability of generating a path $y=(y_0,\dots,y_L)$ from $x$, shooting forward and backward from the $i$-th index, is then
\begin{equation}
\label{generation_shooting}
{\cal P}(x,y) = \prod_{j=0}^{i-1} {\rm \bar{p}}(y_{j+1},y_j) \prod_{j=i+1}^L {\rm p}(y_{j-1},y_j). 
\end{equation}
Notice that the previous path $x$ is present only through the term $y_i = x_i$.
It then follows
\[
r(x,y) = \min \left ( 1, \hA(y_0) \hB(y_L) c_{\rm exact}(x,y) \right ),
\]
with 
\begin{equation}
\label{c_exact}
c_{\rm exact}(x,y) = \frac{\rho(y_0)}{\rho(x_0)} \prod_{j=0}^{i-1} \frac{{\rm p}(y_j,y_{j+1})}{{\rm \bar{p}}(y_{j+1},y_j)}\frac{{\rm \bar{p}}(x_{j+1},x_j)}{{\rm p}(x_j,x_{j+1})} .
\end{equation}

It is clear that, for reasonable discretizations, $P^2(x,y) > 0$ for all paths $x, y$ 
%
%
of positive probability (under mild assumptions on the potential) so
that the correponding Markov chain is irreducible. Since the
measure~(\ref{canonical_path_discretized}) is left invariant by the
dynamics (this is a classical property of Metropolis-Hastings scheme), the corresponding Markov chain is ergodic~\cite{MT}. Notice also that it is enough to consider only the forward or the backward integration steps for the ergodicity to hold, as long as both have a positive probability to occur (and that the possible asymmetry in the corresponding probabilities is accounted for).

In some cases, the microscopic reversibility ratio 
\[
R_{\rm rev}(y_i,y_{i+1}) = \frac{\rho(y_i) \, {\rm p}(y_i, y_{i+1})}{\rho(y_{i+1}) \, {\rm {\bar{p}}}(y_{i+1}, y_i)}
\]
is close to 1, so that $c_{\rm exact}(x,y) \simeq 1$ and the
acceptance/rejection step is greatly simplified. However, this
assumption should always be checked carefully using some preliminary runs since
it is sometimes the case that, even if the reversibility ratio $R_{\rm
  rev}$ is close to 1 pointwise (with a good approximation), it may be
false that $c_{\rm exact}(x,y) \simeq 1$ along the path, especially if
the paths are long (see~\cite{stoltz_PhD} for a more systematic study of
this point). 

\subsection{The brownian tube proposal function}
\label{sec:tube}

A path can also be characterized uniquely by the initial point $x_0$ and
the realization of the
brownian process $W_t$ in~(\ref{SDE}). When discretized, the paths are
then uniquely determined by the sequence of standard gaussian random vectors $U=(U_0,\dots,U_{L-1})$ used to generate the trajectories
using~(\ref{splitting}) (or any discretization of another SDE). This was
already noted in~\cite{CC01}, where a new trajectory was proposed
selecting an index at random and changing only the gaussian random
number associated with this index. 

Since the trajectory is continuous with respect to the realizations of
the brownian motion, any convenient small perturbation of the sequence
of random numbers is expected to generate a path close to the initial
path. Still denoting by ${\rm p}(x_i,x_{i+1})$ the probability to generate a
point $x_{i+1}$ in phase-space starting from $x_i$, using the gaussian random
vectors $U_i$ and $\bar{U}_i$ obtained from standard gaussian random vectors~$G_i$ and~$\bar{G}_i$, the transition probabilities for all classical discretizations
we consider can be writtten as
\[
{\rm p}(x_i,x_{i+1}) = Z^{-1} \exp \left ( - \frac{1}{2} G_i^T \Gamma G_i
\right ), \quad 
\bar{{\rm p}}_{\rm TR}(x_{i+1},x_{i}) = Z^{-1} \exp \left ( - \frac{1}{2} \bar{G}_i^T \Gamma \bar{G}_i \right )
\]
where $Z$ is a normalization constant. In the case of the
discretization~(\ref{splitting}) of the Langevin equation for example,
$\Gamma = V^T V$ where the matrix $V$ allows to recast the correlated
gaussian random vectors $U_i=(U_{1,i},U_{2,i})$ (or $\bar{U}_i$) as standard and
independent gaussian random numbers $G_i$ (or $\bar{G_i}$) through the
transformation $U_i=VG_i$ (or $\bar{U}_i=V \bar{G}_i$) with (see Eq.~(\ref{splitting_conditional}))
\[
V= \left (
\begin{array}{cc}
\dps \sigma_1^{-1} \, {\rm Id}_{dN} & 0\\
\dps \frac{c_{12}}{\sigma_1 \sqrt{1-c_{12}^2}} \, {\rm Id}_{dN} &
\dps \frac{1}{\sigma_2 \sqrt{1-c_{12}^2}} \, {\rm Id}_{dN} \\
\end{array}
\right ).
\]
The idea is then to modify the standard gaussian vectors $G_i$ by an amount $0 \leq \alpha_i \leq 1$ as
\begin{equation}
\label{correlated_gaussian}
\tilde{G}_i = \alpha_i G_i + \sqrt{1 - \alpha_i^2} R_i
\end{equation}
where $R_i$ is a $2dN$-dimensional standard gaussian random vector. A fraction
$\alpha_i$ is associated with each configuration $x_i$ along the path. The
usual shooting dynamics is recovered with $\alpha_i = 0$ for all $i$
(all the Brownian increments are uncorrelated with respect to the
Brownian increments of the modified path), whereas the so-called 'noise
history' algorithm proposed in~\cite{CC01} corresponds to $\alpha_i = 0$
for all $i$ but one $i_0$ for which $\alpha_{i_0}=1$ (in this case, all
the Brownian increments but one are re-used).

The dynamics we propose looks like the shooting dynamics: first, a
position $0 \leq k \leq L$ along the path is chosen at random; a
coefficient $\alpha_i$ is then associated to each configuration along
the path, and a random gaussian vector is proposed starting from the
previous one using~(\ref{correlated_gaussian}); finally, the
corresponding trajectory is integrated forward from the $k$-th
configuration to the $L$-th, and time-reversed from the $k$-th to the
first, and an acceptance/rejection step is done according
to~(\ref{MH_AR}).

It only remains to precise the proposition function ${\cal P}(x,y)$. Denoting
by $(\bar{G}_i^x)_{0 \leq i \leq k-1}, \ (G_i^x)_{k \leq i \leq L-1}$
the standard random gaussian vectors associated with the path $x$ (the
first ones arise from the time reversed integration, the last ones from
a usual foward integration), it follows
\[
{\cal P}(x,y) = w_k \prod_{0 \leq i \leq k-1} {\rm p}_{\alpha_i}(\bar{G}^x_i,
\bar{G}^y_i) \ \prod_{k \leq i \leq L-1} {\rm p}_{\alpha_i}(G^x_i, G^y_i),
\]
where $w_k$ denotes the probability to choose $k$ as a shooting index, and
\[
{\rm p}_\alpha(G,\tilde{G}) = \left ( \frac{1}{\sqrt{2\pi(1-\alpha^2)}}
\right)^d \exp \left ( -\frac{(\tilde{G}-\alpha G)^T (\tilde{G}-\alpha G)}{2(1-\alpha^2)} \right ).
\]

A tuning of the coefficients $\alpha_i$ can then be performed in order
to get the best trade-off between acceptance (which tends to 1 in the
limit $\alpha_i=1$ for all $i$) and decorrelation (which arises in the
limit $\alpha_i \to 0$). An interesting idea could be that $\alpha$ has
to be close to 1 in regions where the generating moves have a chaotic behavior  
(in the sense that even small perturbations to a path
lead to large changes to this path), and could be smaller in regions where the
generating moves have less impact on the paths (so as to increase the decorrelation). From a more practical point of view, possible approaches to obtain such a trade-off are to propose a functional form for the coefficients $\alpha_i$ and to perform short computations to optimize the parameters with respect to some objective function. Some simple choices for the form of the coefficients $\alpha_i$, involving only one parameter (so that the optimization is procedure is easier), are:
\begin{itemize}
\item constant coefficients $\alpha_i = \alpha$;
\item set $\alpha_i = 1$ far from the shooting index, and $\alpha_i$ close to 0 near the shooting index. This can be done by considering $\alpha_i = \min (1, K|i-k|)$ for some $K \geq 0$. 
\end{itemize}
From our experience, the efficiency is robust enough with respect to the choice of the decorrelation coefficients $\alpha_i$.
Notice also that the second functional form allows to recover both the usual shooting and the noise-history algorithm, respectively in the regimes $K \to 0$ and $K \geq 1$. It is therefore expected that, optimizing the efficiency with respect to $K \in [0,1]$, both the shooting algorithm and the noise-history algorithm should be outperformed.

\subsection{Intrinsic measure of efficiency}
\label{sec:local_sampling}

Our aim here is to propose some abstract measure of decorrelation
between the paths, so as to measure some diffusion in path space.
This approach complements the convergence tests based on some observable of interest for the system. We refer to~\cite{DBG02} for some examples of relevant quantities to monitor (and applications to path sampling with deterministic dynamics), and to section~\ref{sec:num_eq} for some numerical results for stochastic dynamics.  

The intrinsic decorrelation is related to the existence of some distance
or norm on path space. Given a distance function ${\rm d}(x,y)$, the quantity
\[
D_p(n) = \left ( \int \int \left [ {\rm d}(y,x) \right ]^p \, P^n(x,dy) \, d\pi(x) \right )^{1/p}
\]
(with $p \geq 1$) precises the {\em average} amount of decorrelation
with respect to the distance ${\rm d}$ for the measure $\pi$ on the path
ensemble. Notice that two averages are taken: one over the initial paths
$x$, and another over all the realizations of the Monte Carlo iterations
starting from $x$ ({\it i.e.} over all the possible end paths $y$,
weighted by the probability to end up in $y$ starting from $x$). 
In practice, assuming ergodicity, $D_p(n)$ is computed as
\[
D_p(n) = \lim_{N \to + \infty} \left ( \frac1N \sum_{k=1}^N {\rm d}^p(x^{k+n},x^k) \right )^{1/p}.
\]
Usual choices for $p$ are $p=1$ or
$p=2$. This last case is considered in~\cite{ceperley95} since a
diffusive behavior over the space is expected with stochastic dynamics,
the most efficient algorithms having the largest diffusion constants $\lim_{n \to +\infty} \sqrt{D_2(n)/n}$.

It then only remains to precise the distance ${\rm d}$, which depends on
the system of interest. Some simple choices are to
\begin{itemize}
\item consider a (weighted) norm $||\cdot||$ on the whole underlying phase-space
  (for position or position/momenta variables) and set
\[
{\rm d}(x,y) = \left ( \frac{1}{L} \sum_{i=0}^{L} \omega_i ||x_i-y_i||^{p'} \right )^{1/p'}
\]
with $p' \geq 1$;
\item consider only a projection of the configurations onto some
  submanifold, such as the level sets of a given (not necessarily
  completely relevant) reaction coordinate or order parameter $\xi$:
\[
{\rm d}(x,y) = \left ( \frac{1}{L} \sum_{i=0}^{L}  \omega_i |\xi(x_i)-\xi(y_i)|^{p'} \right )^{1/p'},
\]
with $p' \geq 1$.
\item align the paths projected onto some
  submanifold around a given value of the reaction coordinate $\xi$:
\begin{equation}
\label{dist_align}
{\rm d}(x,y) = \left ( \frac{1}{2K+1} \sum_{i=-K}^{K} \omega_i |\xi(x_{I+i})-\xi(y_{J+i})|^{p'} \right )^{1/p'},
\end{equation}
with $p' \geq 1$, and $I,J$ such that $\xi(x_I) = \xi(x_J) = \xi^*$
where $\xi^*$ is fixed in advance (for example, if $A$ is characterized
by $\xi=0$ and $B$ by $\xi=1$, then $\xi^*$ could be 1/2). The integer
$K$ represents some maximal window frame so that the distance is really
restricted to a region around the expected or assumed transition point.
In the case when $J-K, I-K < 0$ or $J+K, I+K > L$, the sum is accordingly 
restricted to less than $2K+1$ points. 
\end{itemize}
The weights $\omega_i$ should be non-negative in all cases.

A reasonable choice for non-trivial systems is for example to
use~(\ref{dist_align}) with $p'=1$ and $\omega_i = 1$. This approach ensures that the decorrelations
arising in the initial and final basins $A$ and $B$ are discarded, and
that only the decorrelation arising near the transition region are
important. In this sense, we term this decorrelation as 'local
decorrelation' since we measure how different the transition mechanisms
are. As a measure of 'global decorrelation', we will consider the
transition times. A numerical study based on those lines is presented in section~\ref{sec:num_eq}.

\subsection{Numerical results}
\label{sec:num_eq}

\begin{figure}
\centering
\hspace{-0cm}
\includegraphics[width=7cm]{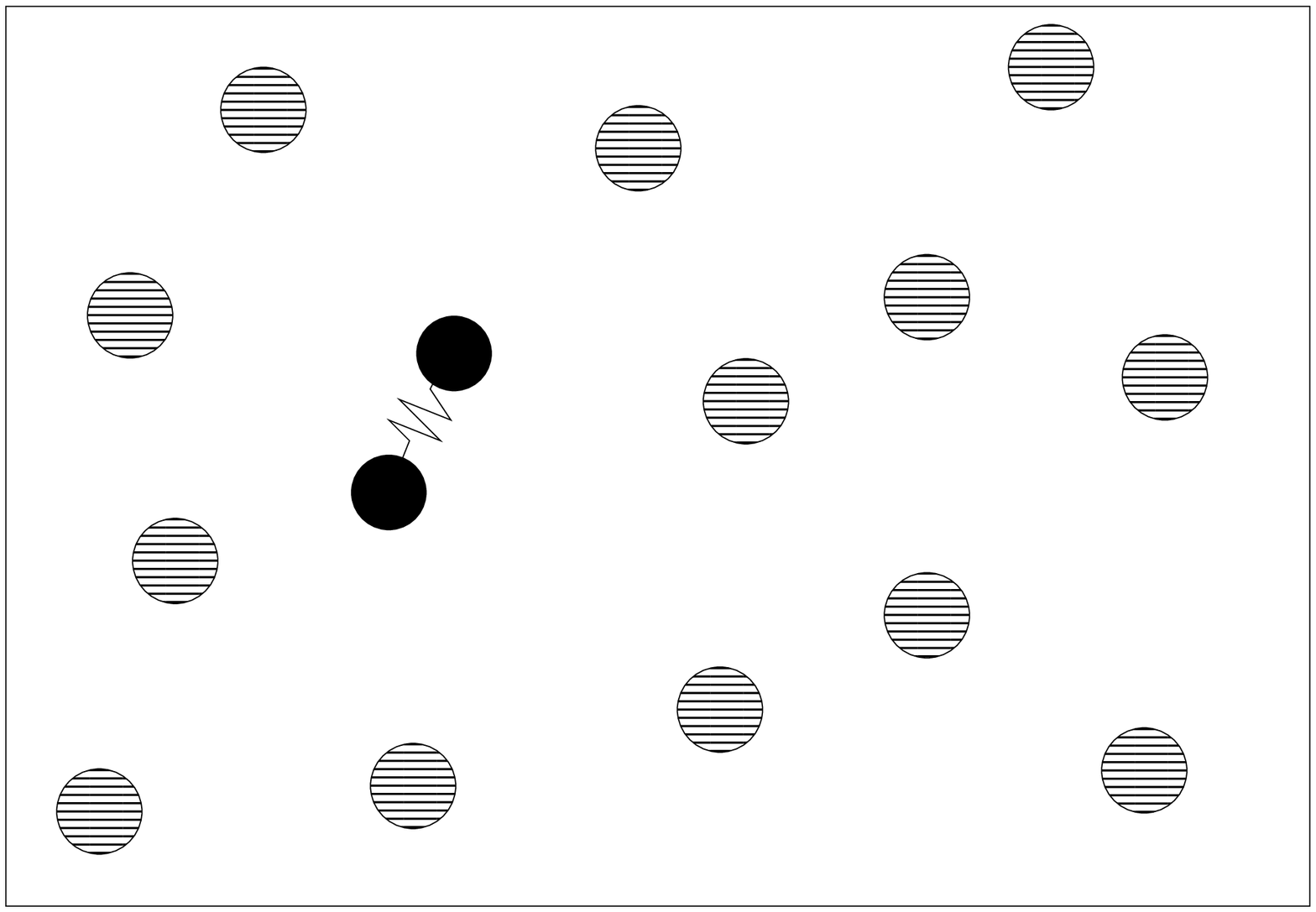}\hfill
\includegraphics[width=7cm]{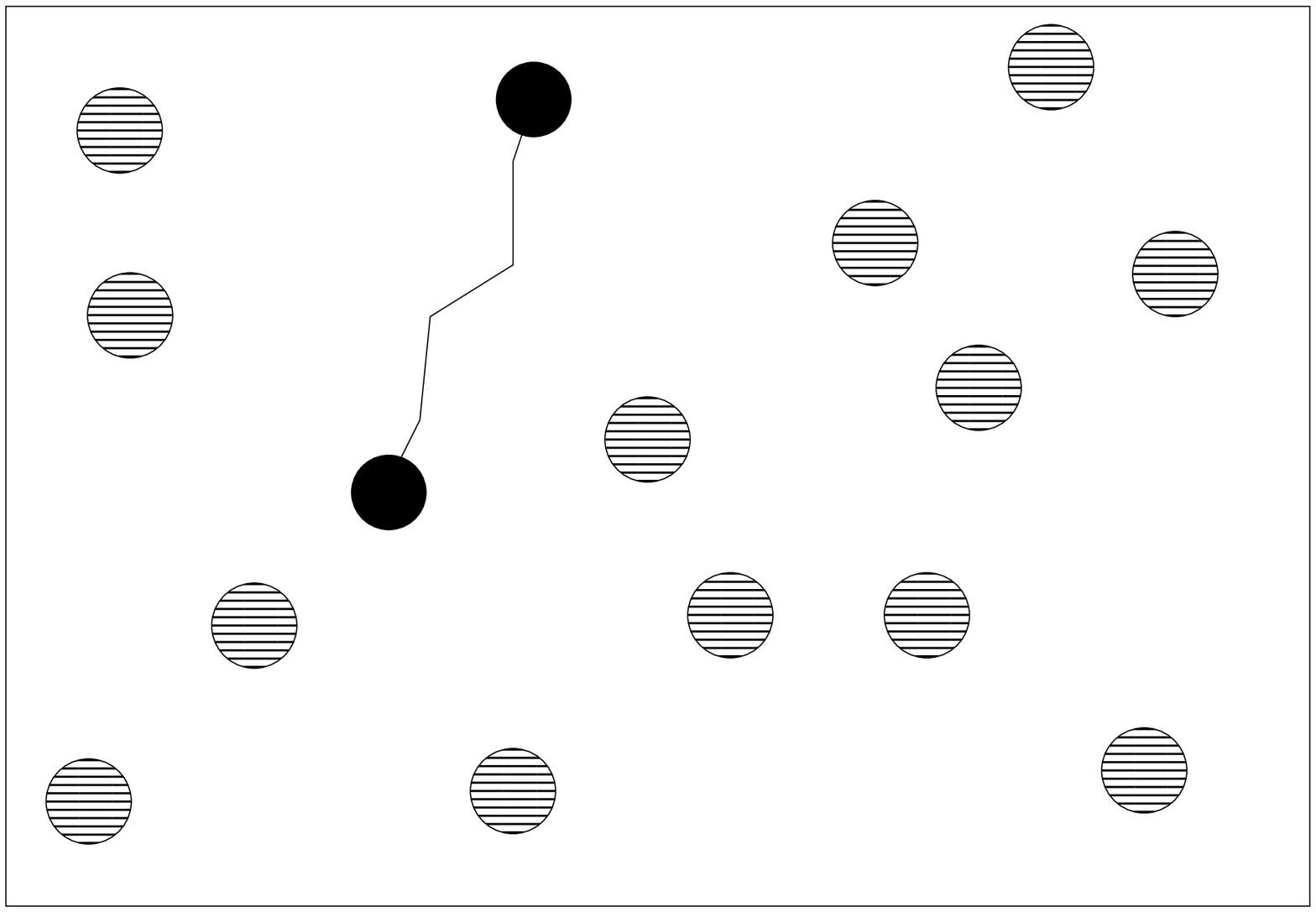}
\caption{\label{WCA_fig} Schematic views of the system, when the
  diatomic molecule is in the compact state (Left), and in the stretched
  state (Right). The interaction of the atoms forming the molecule is
  described by a double well potential, all the other interactions are
  of WCA form.}
\end{figure}

We test the different proposal functions on a model system of conformational
changes influenced by solvation. We consider a system composed of $N$
particles in a periodic box of side length $l_0$, interacting through
the purely repulsive WCA pair potential~\cite{DBC99,SBB88}:
\[
V_{\rm WCA}(r) = \left \{ \begin{array}{cl}
\dps 4 \epsilon \left [ \left ( \frac{\sigma}{r} \right )^{12} 
- \left ( \frac{\sigma}{r}\right )^6 \right ] + \epsilon & {\rm if \ } r \leq r_0, \\
0 & {\rm if \ } r > r_0,
\end{array} \right.
\]
where $r$ denotes the distance between two particles, $\epsilon$ and $\sigma$ are two positive parameters and $r_0=2^{1/6}\sigma$.
Among these particles, two (labeled 1 and 2 in the following) are
designated to form a dimer while the remaining particles are
solvent particles. Instead of the above WCA
potential, the interaction potential between the particles in the dimer 
is a double-well potential
\[
V_{\rm DW}(r) = h \left [ 1 - \frac{(r-r_0-w)^2}{w^2} \right ]^2,
\]
where $h$ and $w$ are two positive parameters. The potential $V_{\rm DW}$ exhibits two energy minima, one corresponding
to the compact state where the bond length of the solute dimer is
$r=r_0$, and one corresponding to the stretched state where the bond
length of the solute dimer is $r(q)=r_0+2w$. The energy barrier
separating both states is $h$. Figure~\ref{WCA_fig} presents a schematic
view of the system. 

We consider the distance~(\ref{dist_align}) for reactive paths ($\pi \equiv \pi_\AB$ in this case), using $p=p'=1$ and $\omega_i = 1$, $\xi(q)
= |q_1-q_2|$, $\xi^* = r_0 + w$. We use the parameters
$L=500 \, \Delta t$, $\beta=1$, $N=16$ particles of masses $1$, $l_0=1.3$, $\sigma=1$,
$\epsilon=1$, $w=0.5$, $\Delta t = 0.0025$, with the sets $A=\{ \xi(q)
\leq r_0+0.6 w\}, B=\{ \xi(q)\geq r_B=r_0+1.4 w \}$ and
averaging over a total of $n=5 \times 10^4$ Monte Carlo moves. We set $K=30$ since
the typical length of the transitions is about 60 time steps with the
parameters used here.

We also consider the correlation in the transition times. We denote by
$\tau(x)$ the transition index of some path $x$. Here, those indexes $\tau$ are
such that $\xi(q_{\tau \Delta t}) = \xi^*$. The correlation function for this
observable is therefore, in the case of reactive paths,
\[
C(n) = \frac{ \dps \int \int (\tau(y) - \langle \tau \rangle_{\pi_\AB})(\tau(x) - \langle \tau \rangle_{\pi_\AB}) \, P^n(x,dy) \, d\pi_\AB(x)}
{\dps \int (\tau(x) - \langle \tau \rangle_{\pi_\AB})^2 \, d\pi_\AB(x)}, 
\]
with $\langle \tau \rangle_{\pi_\AB} = \int \tau(x) d\pi_\AB(x)$
This observable is in some sense complementary to the measure of decorrelation in the transition zone defined above since it measures some global spatial decorrelation of the paths.
In practice, assuming ergodicity, $C$ is approximated as
\[
C(n) = \lim_{N \to +\infty} \frac{\dps \frac1N \sum_{k=1}^N 
\tau(x^{n+k})\tau(x^k) - \left ( \frac1N \sum_{k=1}^N 
\tau(x^{n+k}) \right )\left ( \frac1N \sum_{k=1}^N 
\tau(x^{k}) \right )}
{\dps \frac1N \sum_{k=1}^N \tau(x^k)^2 - \left ( \frac1N \sum_{k=1}^N 
\tau(x^{k}) \right )^2}.
\]

\begin{figure}[h]
\begin{center}
\includegraphics[width=7cm]{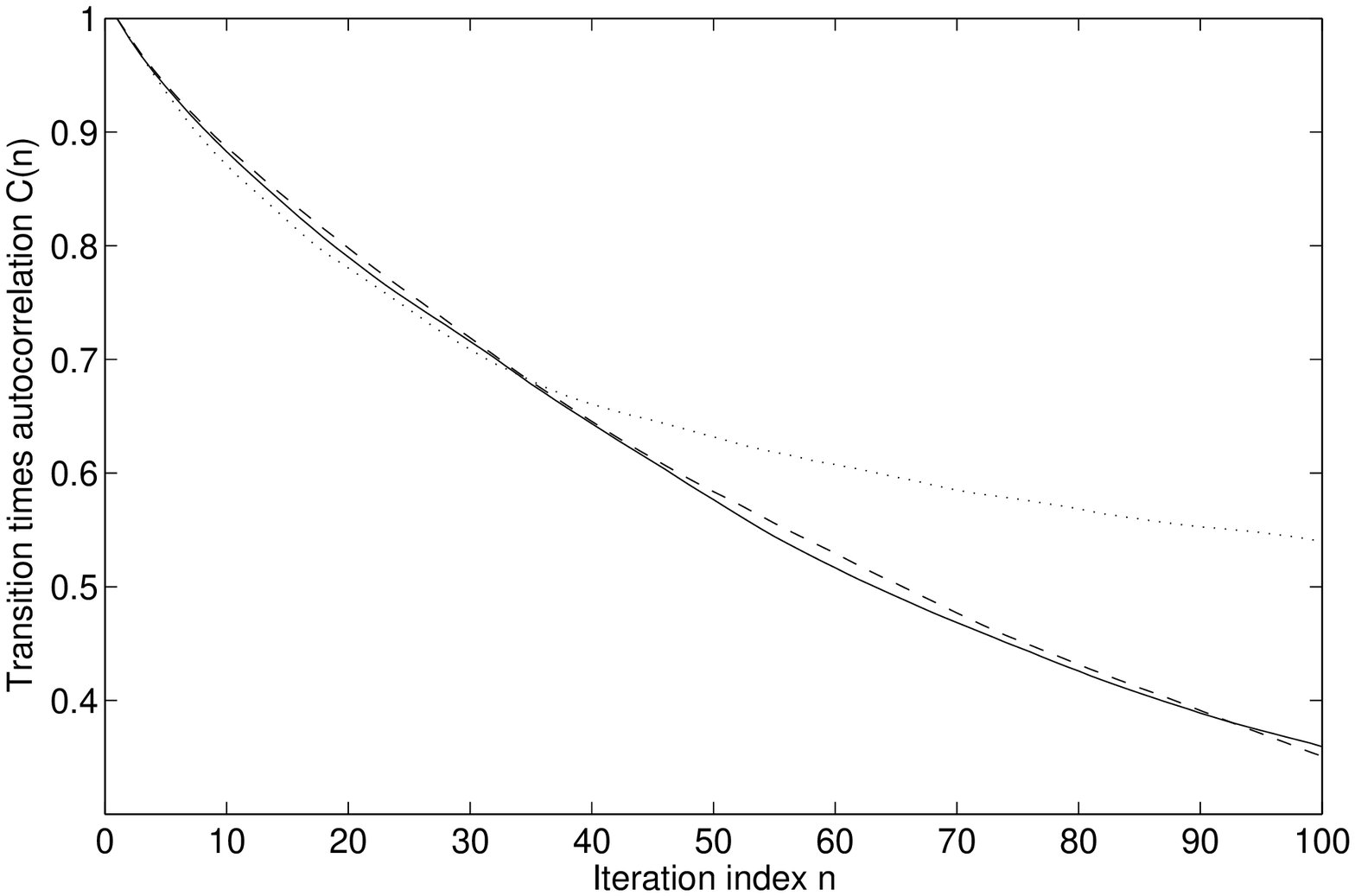}
\includegraphics[width=7cm]{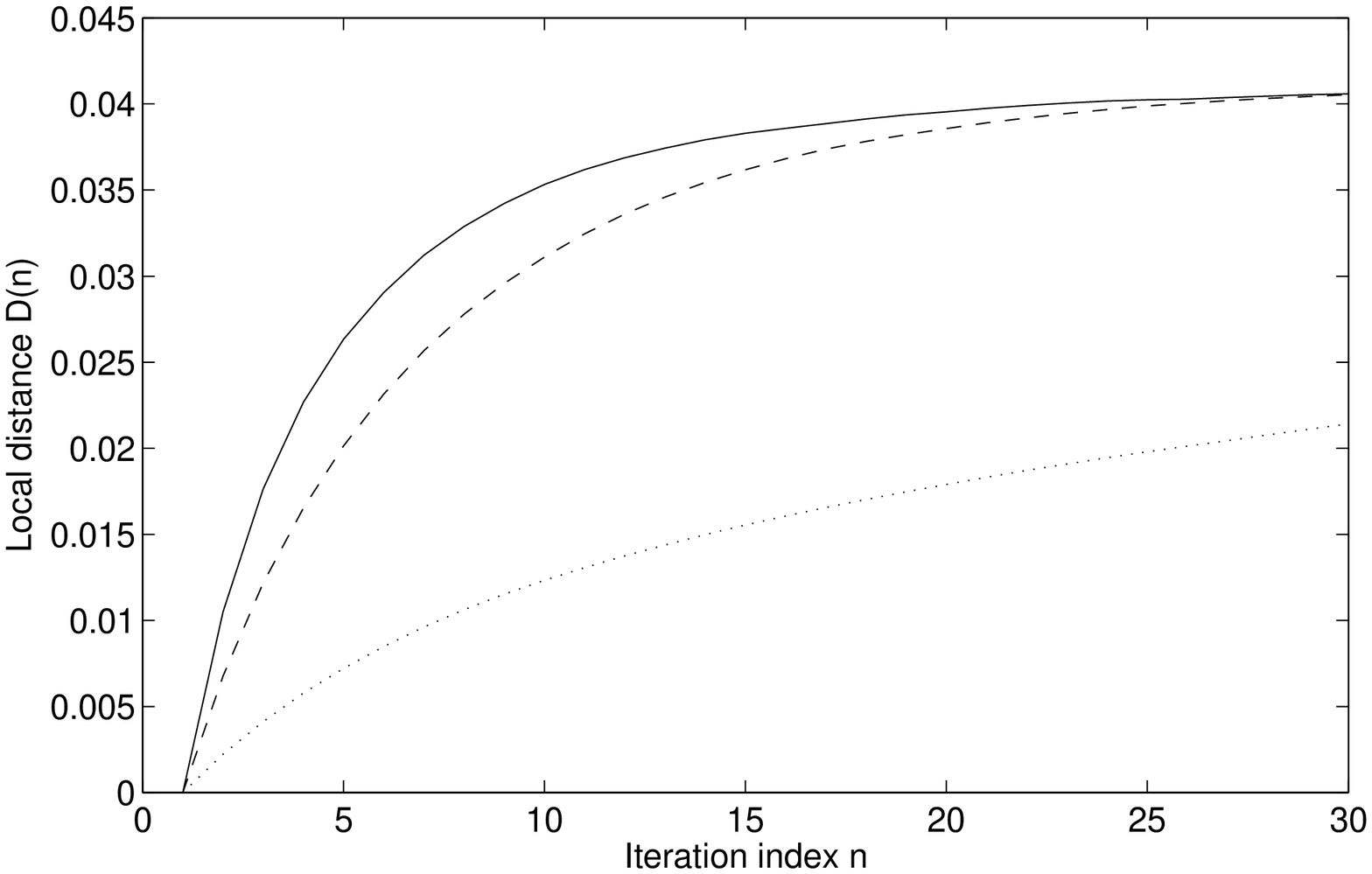}
\caption{\label{comp_efficiency} Comparison of efficiencies for
different Metropolis-Hastings proposal moves for $h=5$. Left:
Plot of the correlation of the transition times $C(n)$ (related to some global sampling efficiency). Right: Plot
of $D(n)$ (local sampling efficiency) for the brownian tube proposal with $\alpha \equiv 0.8$ (solid line), usual shooting dynamics (dashed line), and noise history (dotted line).}.
\end{center}
\end{figure}

\begin{figure}[h]
\begin{center}
\includegraphics[width=7cm]{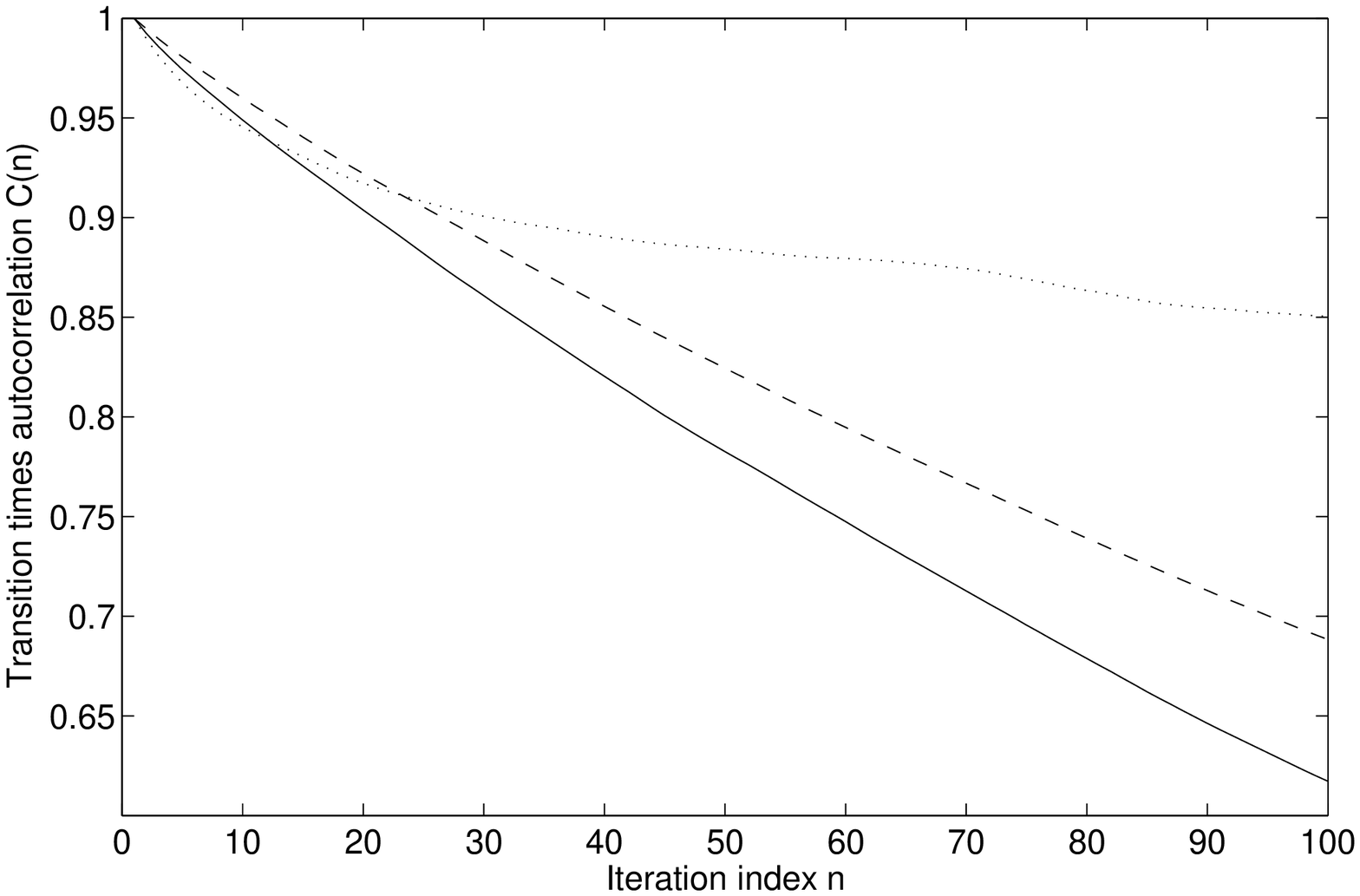}
\includegraphics[width=7cm]{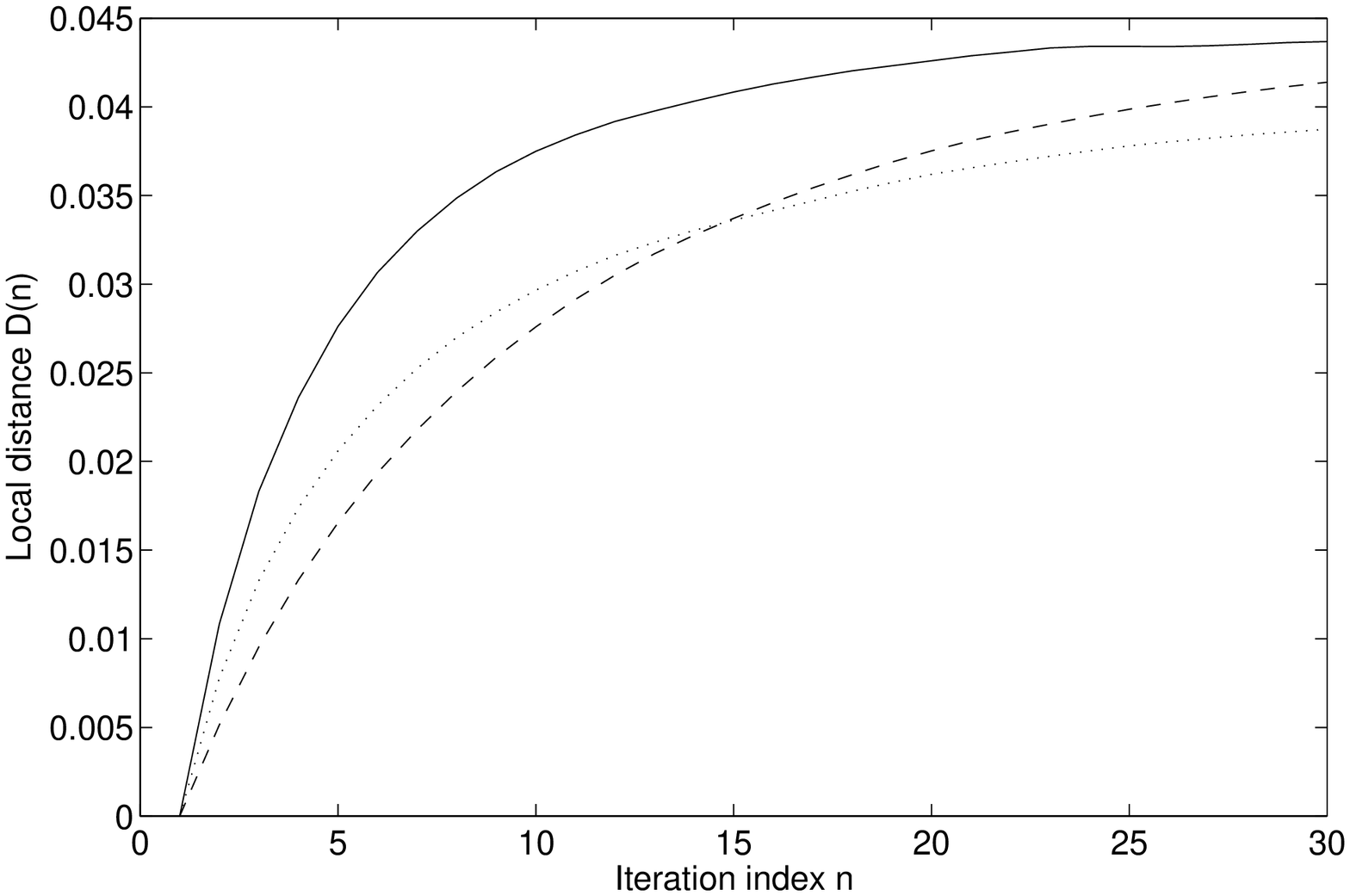}
\caption{\label{comp_efficiency2} Comparison of efficiencies for
  different Metropolis-Hastings proposal moves for $h=10$.}.
\end{center}
\end{figure}

\begin{figure}[h]
\begin{center}
\includegraphics[width=7cm]{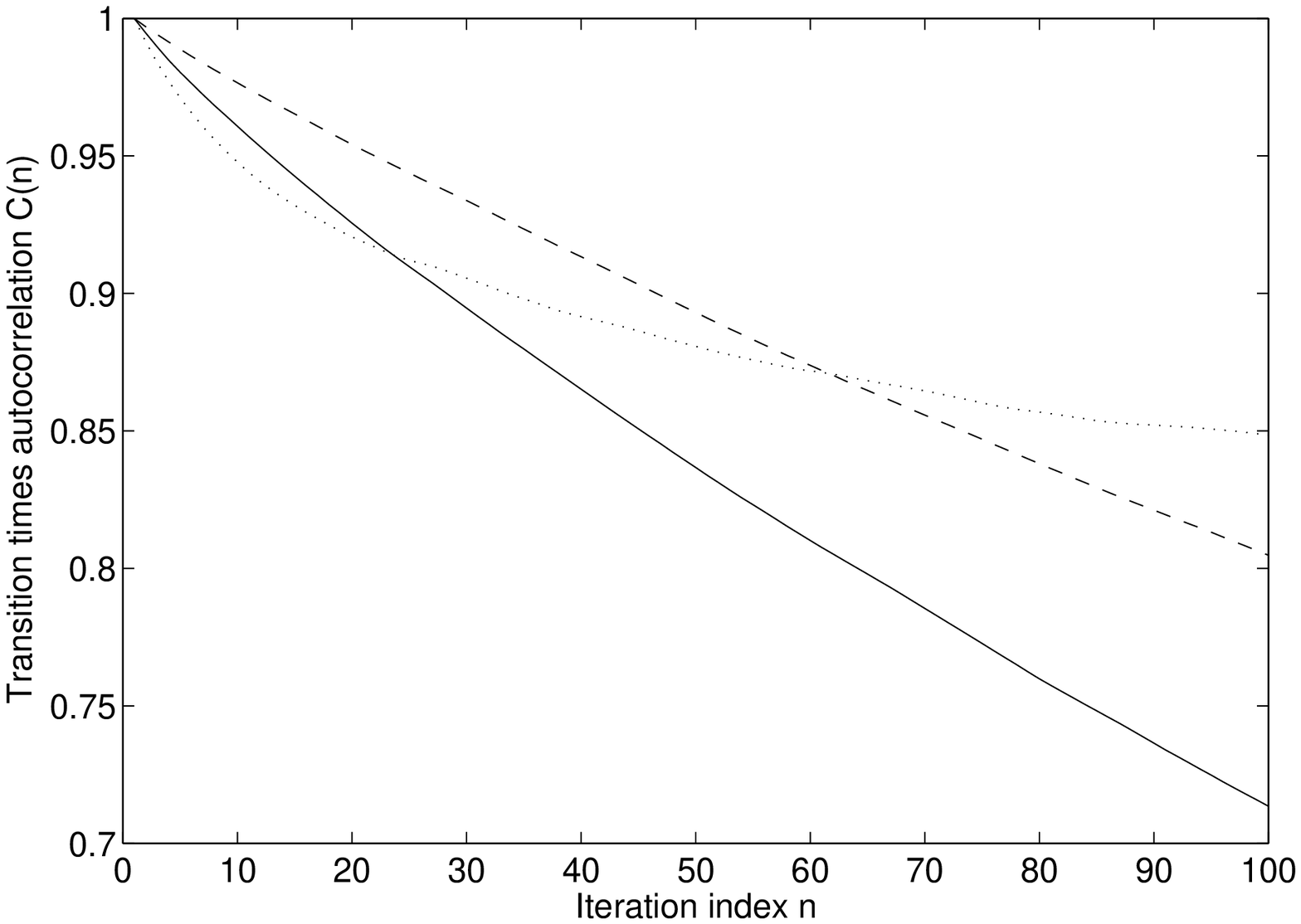}
\includegraphics[width=7cm]{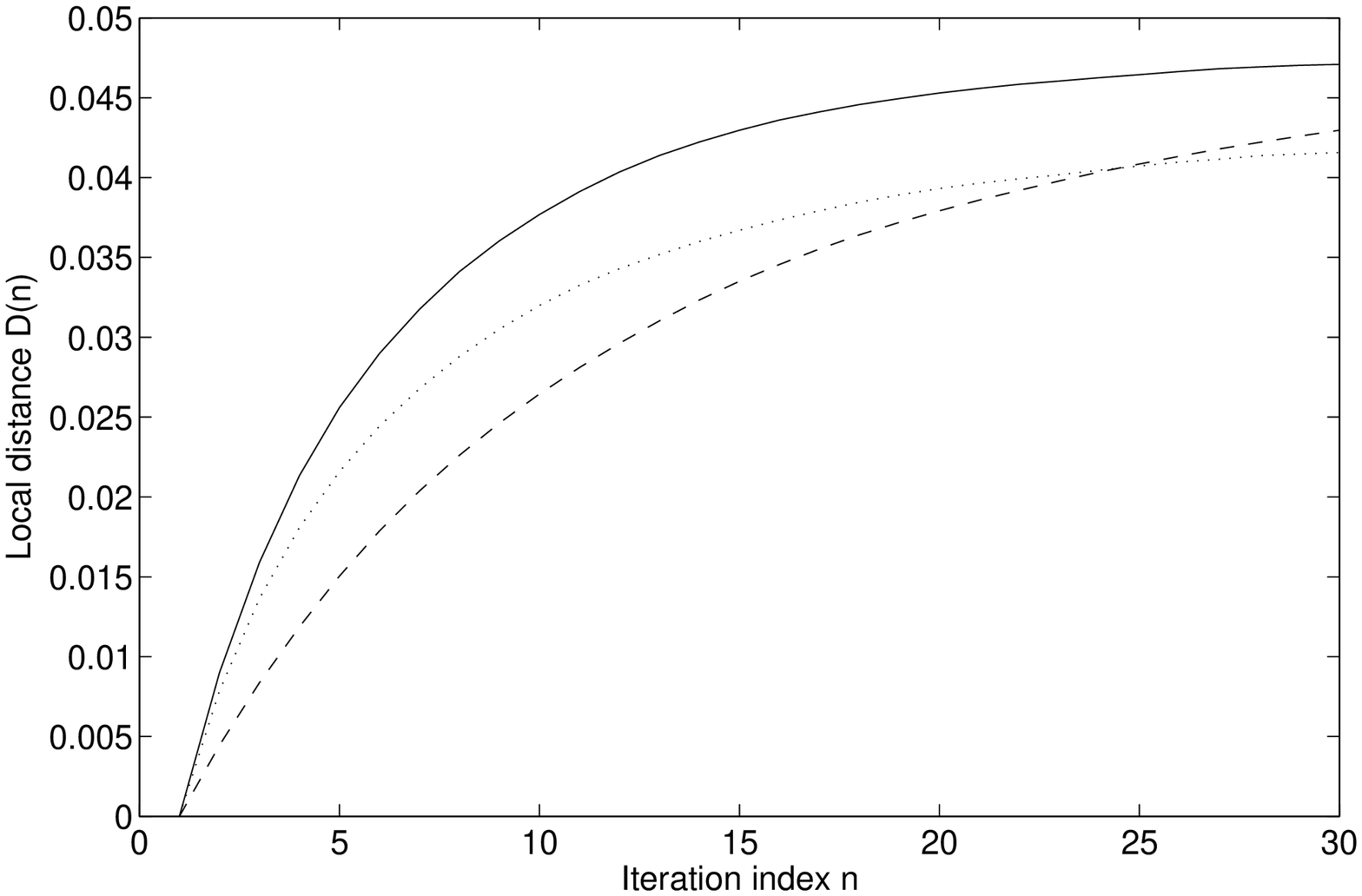}
\caption{\label{comp_efficiency3} Comparison of efficiencies for
  different Metropolis-Hastings proposal moves for $h=15$.}.
\end{center}
\end{figure}

Figures~\ref{comp_efficiency} to~\ref{comp_efficiency3} present some
plots of $D(n)$ and $C(n)$ for $h=5,10,15$,
for the usual shooting dynamics, the noise-history algorithm, and the
brownian tube proposal (with $\alpha_i = 0.8$ for all $i$). The average
acceptance rates are also presented in Table~\ref{tab:rejection}.
Notice that no shifting moves~\cite{DBG02} are used in order to compare the intrinsic efficiencies of the proposal functions. It is likely that these moves would help improving the decorrelation rate of the sampling.

\begin{table}
\centering
\begin{tabular}{|c | c c c |}
\hline
$h$ & 5 & 10 & 15 \\
\hline
Shooting & 24.4 & 18.1 & 15.2 \\
Noise history & 96.7 & 85.7 & 81.2 \\
Brownian tube ($\alpha_i = 0.8$) & 47.2 & 48.1 & 33.0 \\
\hline
\end{tabular}
\caption{\label{tab:rejection} Acception rate ($\%$) as a function of
  $h$ for the three proposal functions considered.}
\end{table}

For the shooting algorithm, many paths are rejected so that the local
decorrelation (measured by $D(n)$) is rather poor, especially at short algorithmic times 
and for high barriers (in any cases, lower than for the brownian tube proposal).
But when a path is
accepted, it is already very decorrelated from the previous one, so that
the global decorrelation (measured by $C(n)$) is indeed decreasing
rapidly enough. For the noise-history algorithm, the picture is somewhat
inverted: since the acceptance rate is very high, even for high barriers, the local
decorrelation is quite efficient, but the global decorrelation is not
since small local changes make it difficult to change the global
features of the paths. The brownian tube approach tries to balance the
local and global decorrelations. This is also reflected by a more balanced
acceptance/rejection rate.

In conclusion, the brownian
tube proposal with the above correlation function is the most efficient
sampling scheme in the case considered here. The efficiency could be further increased
by a more systematic tuning of the parameters of the correlation factors
$\alpha_i$, possibly depending on the shooting index $k$. In general,
since the usual proposal functions are specific cases of the brownian
tube proposal function, it is expected that there is always a parameter
range such that this new algorithm outperforms the previous ones.

%
%

\section{(Non)equilibrium sampling of the path ensemble}
\label{sec:noneq}

The previous section was dealing with equilibrium sampling of
paths. However, when (free) energy barriers in path space are large, direct sampling
of paths can be inefficient, since the existence of metastable path sets may considerably
slow down the numerical convergence. It is therefore appealing to
perform some kind of {\it simulated annealing} on paths.
A regular simulated annealing strategy would be to first sample paths at
a higher temperature, and then to cool the sample to the target
temperature (see~\cite{VS01} for a
simulated tempering version of such an idea).
Reactive paths can also be otained by constraining
progressively the paths to end up in $B$.
This approach also has the nice feature that it does not ask for an
initial guess to start sampling $\pi_\AB$.
Finally, a byproduct of such a switching is the ratio of partition functions in path space
\begin{equation}
\label{cte_reac}
C(L\Delta t) = \frac{Z_\AB(L\Delta t)}{Z_A(L\Delta t)},
\end{equation}
where $Z_A, Z_\AB$ are such that
\[
\pi_{A}(x) = Z_A(L\Delta t)^{-1} \hA(x_0) \rho(x_0) \prod_{i=0}^{L-1} {\rm p}(x_i,x_{i+1}),
\]
and
\[
\pi_{AB}(x) = Z_\AB(L\Delta t)^{-1} \hA(x_0) \rho(x_0) \prod_{i=0}^{L-1} {\rm p}(x_i,x_{i+1}) \hB(x_L)
\]
are probability measures. The function $C$ in~(\ref{cte_reac}) has to be computed at
least once to obtain rate constants in practice~\cite{DBG02}. The
associated free-energy difference in path space is $\Delta
F_{A \to AB}(L\Delta t) = -\ln (C(L\Delta t))$.

We start this section by recalling the extension of the classical switching dynamics for
nonequilibrium dynamics in phase space to nonequilibrium switching
between path ensembles~\cite{GD04}. This method is convenient to compute
free energy differences, but the final sample of paths obtained is
very degenerate. We therefore present in section~\ref{sec:IPS} the application to
path sampling of a birth/death process introduced
in~\cite{Rousset06,RS06}, which allows to keep the sample at equilibrium
at all times during the switching. This equilibration may be important
in some cases to compute the right free energy values~\cite{RS06}, and
allows in any cases to end up with a non-degenerate sample of paths and
reduce the empirical variance.
We will focus in the sequel on switching from constrained to
unconstrained paths, but an extension to simulated annealing (cooling
process) is straightforward.

\subsection{Switching between ensembles of paths}
\label{sec:switching_non_eq}

We present in this section the approach of~\cite{GD04}, where
the switching from unconstrained to constrained path ensembles is done by enforcing progressively
the constraint on the end point of the path over a time interval
$[0,T]$. The constraint is usually parametrized using some order parameter. This order parameter is the same as the one used for usual computation of reaction rates in the TPS framework (and even for
more advanced techniques such as Transition Interface Sampling
(TIS)~\cite{vEMB03,vEB05}). The point is that this approximate order
parameter needs not to be a ``good'' reaction coordinate (or a complete one) since the
general path sampling approach should help to get rid of some problems
arising from a wrong choice of order parameter (see {\it e.g}~\cite{vanErp06} for a recent study on this topic).

Assuming an order parameter is given, 
we can consider a switching schedule $\lambda=(\lambda^0,\dots,\lambda^n)$ such that
$\lambda^0=0$ and $\lambda^n=1$ and a family of functions
$h_{\lambda}$ such that 
\[
h_0 = {\bf 1}, \quad h_1 = {\bf 1}_B.
\] 
We also introduce the family of measures associated with the functions
$h_{\lambda}$:
\begin{equation}
\label{canonical_path_discretized_lambda}
\pi_{\lambda}(x) = Z_{L,\lambda}^{-1} \hA(x_0) \rho(x_0)
\prod_{i=0}^{L-1} {\rm p}(x_i,x_{i+1}) h_{\lambda}(x_L).
\end{equation}
We omit in the sequel the explicit dependence of the partition
functions $Z$ on $L$ and $\Delta t$.
An energy $\cE_\lambda(x)$ can then formally be associated to a path $x$ as
\[
\pi_{\lambda}(x) = Z_{L,\lambda}^{-1} {\rm e}^{- \cE_\lambda(x)}. 
\]
The aim is to sample from $\pi_1 \equiv \pi_\AB$, which is usually a difficult
task, and sometimes not directly feasible. It may be easier to use a
sample of $\pi_0 = \pi_A$ (which is much easier
to obtain), and to transform it through some switching dynamics into a
weighted sample of $\pi_1$. Starting from a path $x^{k,0}$, the weight factor for a resulting path
$x^{k,n}$ is of the form ${\rm e}^{-W^{k,n}}$ where $W^{k,n}$ is the work exerted on an unconstrained path to constrain it
to end in $B$. We now precise the way the work is computed.

Consider an unconstrained initial path $x^0=(x^0_0,\dots,x^0_L)$ sampled
according to $\pi_0$, and a discrete schedule
$(\lambda^0,\dots,\lambda^n)$. The dynamics in path space is as follows:

\begin{algo}[See Ref.~\cite{GD04}]
Starting from $W^0=0$ and $m=0$,
\begin{itemize}
\item Replace $\lambda^m$ by $\lambda^{m+1}$;
\item Update the work as $W^{m+1} = W^m + \cE_{\lambda^{m+1}}(x^m)-\cE_{\lambda^{m}}(x^m)$;
\item Do a Monte Carlo path sampling move using a Metropolis-Hastings
  scheme with the measure $\pi^{\lambda^{m+1}}$ (using for example the
  usual shooting moves with a Langevin dynamics, or the Monte Carlo move
  designed for path switching presented in Appendix~A), so that
  the current path $x^{m}$ is transformed into the new path $x^{m+1}$.
\end{itemize}
\end{algo}

This procedure is repeated for independent initial conditions $x^{k,0}$,
so that a sample of $M$ end paths $(x^{1,n},\dots, x^{M,n})$ with
weights $({\rm e}^{-W^{1,n}},\dots,{\rm e}^{-W^{M,n}})$ is obtained.
Besides, an estimation of the rate constant is given by the exponential average
\[
C_M(L\Delta t) = - \ln \left ( \frac1M \sum_{k=1}^M{\rm e}^{-W^{k,n}} \right ),
\]
and it can be shown that $C_M \to C$ when $M \to + \infty$.

Since the realizations of the switching procedure are independent provided the initial conditions are independent, 
the random variables $\{{\rm e}^{-W^{k,n}}\}_k$ are i.i.d. A confidence interval can be obtained
for $C_M$ as 
\[
C_{M,\sigma_{\rm c}}^- \leq C_M \leq C_{M,\sigma_{\rm c}}^+, \quad \quad {\rm with} \ C_{M,\sigma_{\rm c}}^\pm = - \ln \left ( \frac1M
  \sum_{k=1}^M{\rm e}^{-W^{k,n}} \pm \sigma_{\rm c} \sqrt{\frac{V_M}{M}} \right ),
\]
where the empirical variance is
\[
V_M = \frac{1}{M} \sum_{k=1}^M \left ( {\rm e}^{-W^{k,n}} -
    \frac1M \sum_{l=1}^M{\rm e}^{-W^{l,n}} \right )^2.
\]
A confidence interval on the free energy difference is then
\[
-\ln C_{M,\sigma_{\rm c}}^- \leq \Delta F_{A \to AB} \leq -\ln C_{M,\sigma_{\rm c}}^+.
\]
For example, the 95 $\%$ confidence interval corresponds to $\sigma_{\rm c} = 1.96$.

Of course, it may the case that the variance of the work distribution is large, so
that only very few paths are relevant (and the confidence interval for
the rate constant is large). Therefore, most
computational effort is discarded in the end. A method enabling an on
the fly sorting out of the irrelevant would be an interesting
improvement of the method. Such a procedure could also concentrate the
efforts on important transition tubes. The Interacting Particle Systems
(IPS), already used in the context of nonequilibrium free energy
differences~\cite{RS06}, is such an approach. 

\subsection{Enhancing the number of relevant paths}
\label{sec:IPS}

We present here an extension of a birth/death process, introduced
for equilibrating a simulated annealing process done at finite rate (and
therefore out of equilibrium), to the case of path
sampling. This procedure can be seen as a time continuous resampling,
and avoids the degeneracy of the paths weights (see also the related population Monte-Carlo algorithms~\cite{Iba}). 
The idea of IPS is to switch several paths in parallel, and to
attach exponentially distributed birth and death times to each
path. The death time of the path is decreased when the work
exerted on it is higher than the average work; when this time is zero, a
new exponentially distributed death time is generated,
the path is suppressed, and replaced by another path picked up at random
among the other paths. The birth time of the path is decreased when the work
exerted on it is lower than the average work; when this time is zero, a
new exponentially distributed birth time is generated,
another path picked up at random is suppressed, and is replaced by the
path giving birth. In all cases (birth or death), the works attached to
a path are kept. We refer to~\cite{RS06} for a proof of the consistency of the method. 

\begin{algo}
Consider an initial distribution $(x^{1,0},\dots,x^{M,0})$ generated from
$\pi_0$. Generate independent times $\tau^{k,b},\tau^{k,d}$ 
from an exponential law of mean $1$. Consider two additional variables 
$\Sigma^{k,b}, \Sigma^{k,d}$ per replica, initialized at $0$.
\begin{itemize}
\item Replace $\lambda^m$ by $\lambda^{m+1}$;
\item Update the works as $W^{k,m+1} = W^{k,m} +\Delta \cE^{k,m} =W^{k,m} 
  \cE_{\lambda^{m+1}}(x^{k,m}) -\cE_{\lambda^{m}}(x^{k,m})$, 
and compute the mean work update $\overline{\Delta \cE}^{m} = M^{-1} \sum_{1 \leq k \leq M} \Delta \cE^{k,m} $;
\item (Diffusion step) Do a Monte Carlo path sampling move using a Metropolis-Hastings
  scheme with the measure $\pi_{\lambda^{m+1}}$, so that $x^{k,m}$ is
  transformed into $x^{k,m+1}$.
\item (Birth/death process) Update the variables $\Sigma^{k,b}$ and $\Sigma^{k,d}$ as
\[
\Sigma^{k,b} = \Sigma^{k,b} + \beta (\overline{\Delta \cE}^m - \Delta \cE^{k,m})^-,
\]
and
\[
\Sigma^{k,d} = \Sigma^{k,b} + \beta (\overline{\Delta \cE}^m - \Delta \cE^{k,m})^+.
\]
(Death) If $\Sigma^{k,d} \geq \tau^{k,d}$, select an index $m \in \{1,\dots,M\}$ at random,  
and replace the $k$-th path by the $m$-th path. Generate a new time
$\tau^{k,d}$ from an exponential law of mean $1$, and set $\Sigma^{k,d} = 0$;

(Birth) If $\Sigma^{k,b} \geq \tau^{k,b}$, select
an index $m \in \{1,\dots,M\}$ at random, and replace the $m$-th path by
the $k$-th path. Generate a new time $\tau^{k,b}$ from an exponential
law of mean $1$, and set $\Sigma^{k,b} = 0$;
\end{itemize}
\end{algo}

Then, each path has weight 1 in the end, and the final sample
$(x^{1,n},\dots, x^{M,n})$ is distributed according to $\pi_1 \equiv
\pi_\AB$.
In this case, an estimation of the rate constant is given by the simple average
\[
C_M(L\Delta t) = \frac1M \sum_{k=1}^M W^{k,n},
\]
and it can be shown that $C_M \to C$ when $M \to + \infty$.
A confidence interval for the free energy difference can be obtained as in section~\ref{sec:switching_non_eq} as
\[
C^{{\rm IPS}, -}_{M,\sigma_{\rm c}} \leq \Delta F_{1 \to AB}  \leq C^{{\rm
  IPS},+}_{M,\sigma_{\rm c}}, \quad \quad {\rm with} \ C^{{\rm IPS}, \pm}_{M,\sigma_{\rm c}} = \frac1M
  \sum_{k=1}^M W^{k,n} \pm \sigma_{\rm c} \sqrt{\frac{V^{\rm IPS}_M}{M}},
\]
the empirical variance being
\[
V^{\rm IPS}_M = \frac{1}{M} \sum_{k=1}^M \left ( W^{k,n} -
    \frac1M \sum_{l=1}^M W^{l,n} \right )^2.
\]

\subsection{Numerical results} 

We compute here the free energy differences while constraining paths to
for the WCA model system introduced in section~\ref{sec:num_eq}. This is
done either with plain nonequilibrium switching, or with the IPS equilibration. 
Let us notice that the energy is fixed in~\cite{GD04} 
while we rather have to fix the temperature in the stochastic setting,
so that a straightforward comparison of the results is not possible. We
set $\beta = 1$ in the sequel.
The other parameters are the same as in~\cite{GD04}: $N=9$ particles, $h=6$,
$\sigma=1$, $\epsilon=1$, the particle density $\rho = 0.6
\sigma^{-2}$, $w=0.25$, and the sets $A=\{ \xi(q) \leq \xi_A = 1.3 \sigma\}, B=\{ \xi(q)\geq
\xi_B = 1.45 \sigma \}$. The trajectory length is $L=320 \, \Delta t$ and
$\Delta t = 0.0025$, so that $L \Delta t = 0.8
(m\sigma^2/\epsilon)^{1/2}$.

We perform a total of $n$ MC moves (using the
brownian tube proposal function (with $\alpha_i = \alpha = 0.8$ for all $0 \leq i \leq L-1$). The function $h_\lambda$ is the one given in~\cite{GD04}:
\[
h_\lambda(q) = {\rm e}^{-\lambda K (1-{\bf 1}_B(q)) (\xi_B-\xi(q))}
\] 
with $K=100$. The switching schedule is $\lambda^i = (i/n)^2$.

\begin{table}
\centering
\begin{tabular}{|c c | c c c |}
\hline
M & n & {\rm Backward} & {\rm Forward} & {\rm IPS (forward)} \\
\hline
2000 & 2000 & 4.83 \ (4.61-5.02) & 5.43 \ (5.28-5.61) & 4.82 \ (4.78-5.85) \\
2000 & 5000 & 5.34 \ (5.04-5.58) & 5.41 \ (5.32-5.50) & 5.19 \ (5.16-5.23)\\
2000 & 10000 & 5.45 \ (5.32-5.58) & 5.40 \ (5.34-5.46) & 5.40 \ (5.36-5.43) \\
2000 & 15000 & 5.42 \ (5.35-5.49) & 5.40 \ (5.35-5.45) & 5.45 \ (5.42-5.48)\\
\hline
\end{tabular}
\caption{\label{tab:df} Free energy differences $\Delta F_{A \to AB}$
  computed for different switching lengths $n$, using a sample
  of $M=2000$ paths. The results are presented under the form "$C_M \ (C_{M,\sigma_{\rm c}}^- - C_{M,\sigma_{\rm c}}^+)$" with $\sigma_{\rm c} = 1.96$ (the value corresponding to a 95 $\%$ confidence interval).}
\end{table}

\begin{figure}[h]
\includegraphics[width=7.3cm]{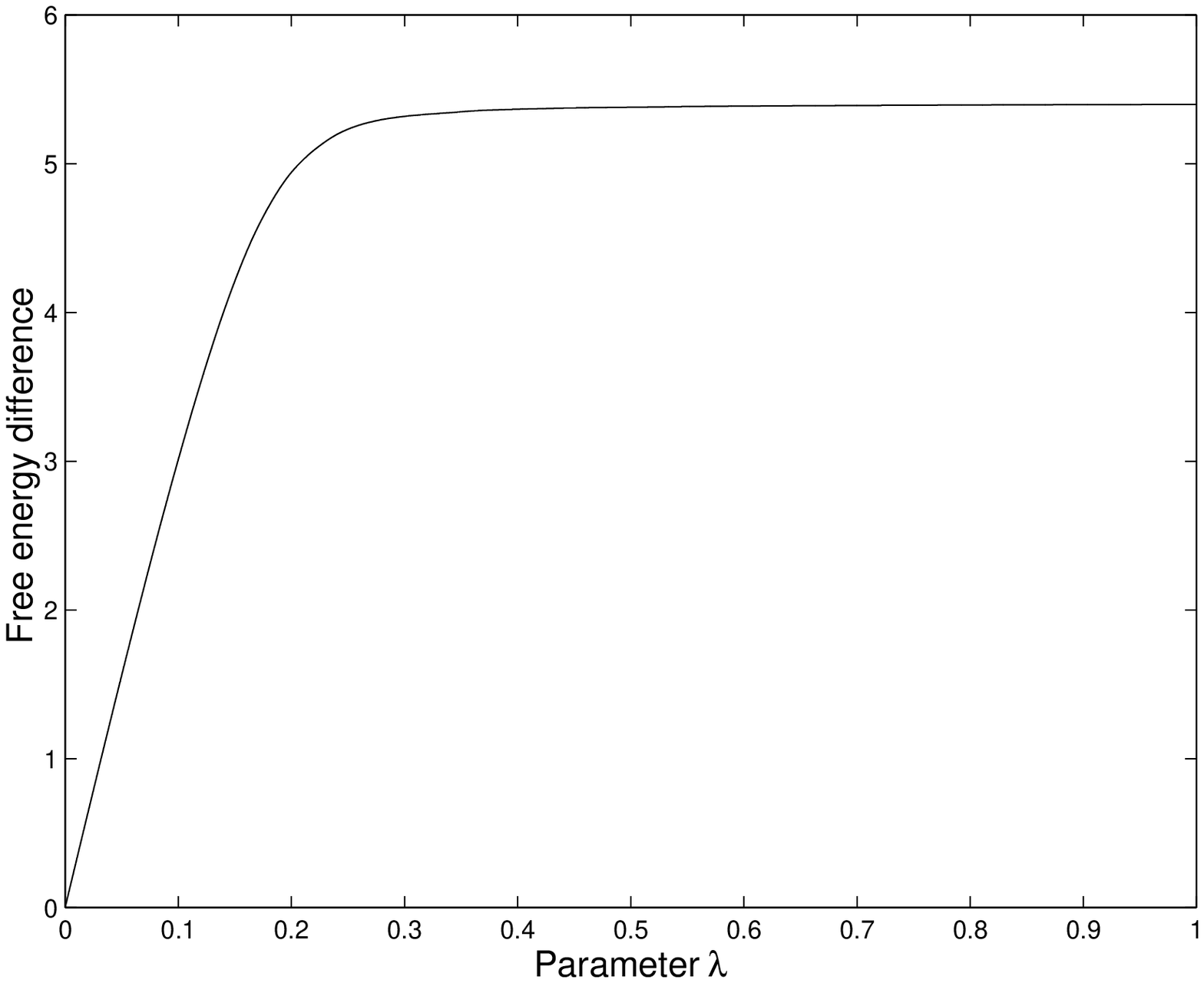}
\includegraphics[width=7.5cm]{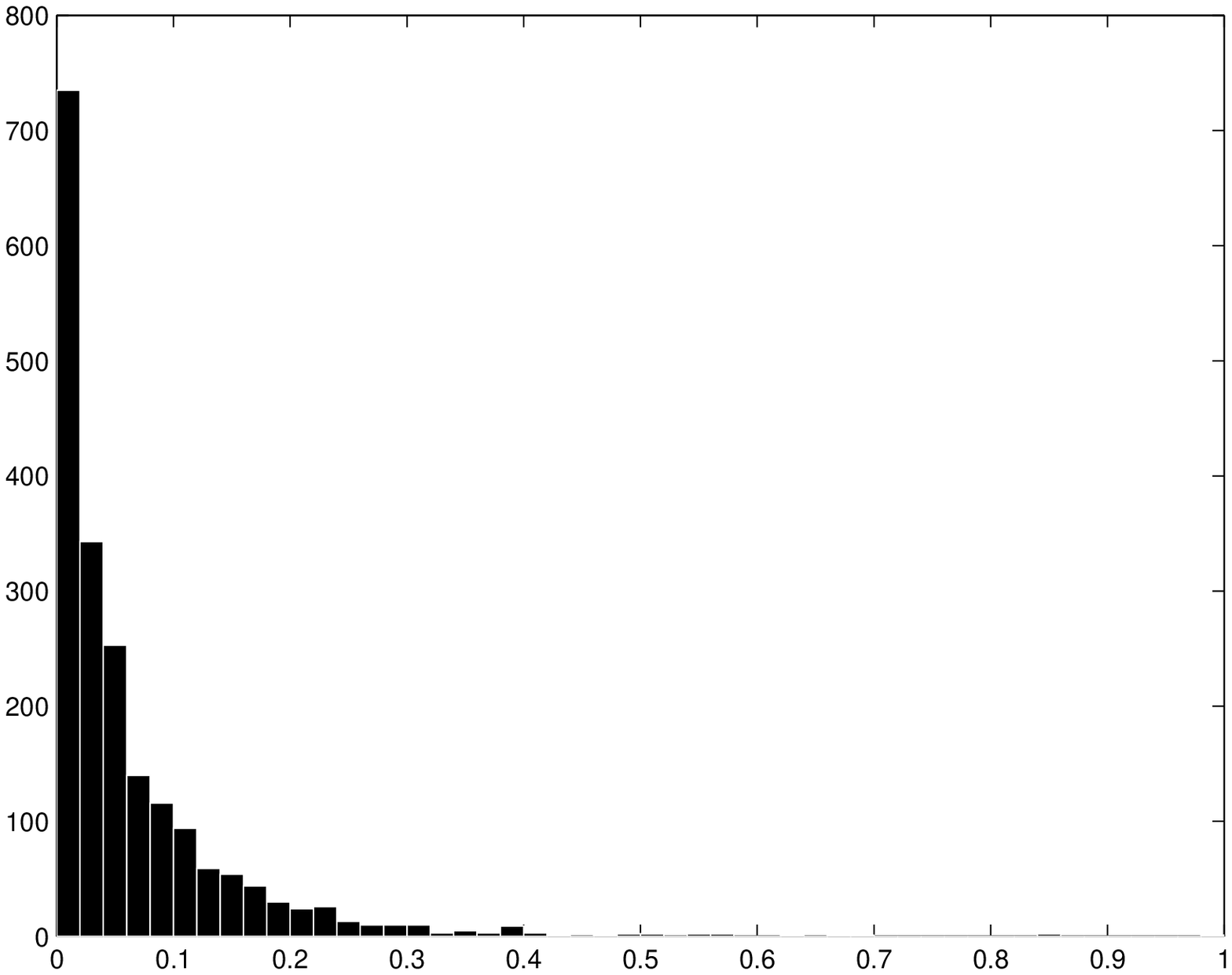}
\caption{\label{fig:df} Left: Free energy profile for a forward switching, computed for $M=2000$ and $n=10^4$, using a plain nonequilibrium switching. Right: Histogram of the weights $w_k$ of the final sample as given by~(\ref{eq:weights}).}
\end{figure}

A typical free energy difference profile is presented in Figure~\ref{fig:df} for $M=2000$ and $n=10000$, as well as the associated weights for the plain nonequilibrium switching. These weights are the Jarzynski weights renormalized by the total weight (in order to define a probability distribution):
\begin{equation}
\label{eq:weights}
w_k = \frac{{\rm e}^{-W^{k,n}}}{\sum_{l=1}^M {\rm e}^{-W^{l,n}}}.
\end{equation}
Notice that the sample is very degenerate since very many paths have negligible weights, and the relevant paths are exponentially rare. Recall also that the paths all have weight 1 with the IPS algorithm.

Some free energy differences are presented in Table~\ref{tab:df} for different values of $n$ (keeping $M$ fixed).
The switchings are slow enough when the confidence intervals for free energy differences computed by constraining paths agree ('forward' switching) overlap with confidence intervals for free energy differences obtained by starting from a sample of constrained paths and removing progressively the constraint ('backward' switching). This is the case here for $n=5000,10000,15000$ (but not when $n=2000$). The results show that IPS agrees with the usual Jarzynski switching, the confidence interval on the results being however lower. 

\begin{figure}[h]
\includegraphics[width=7.5cm]{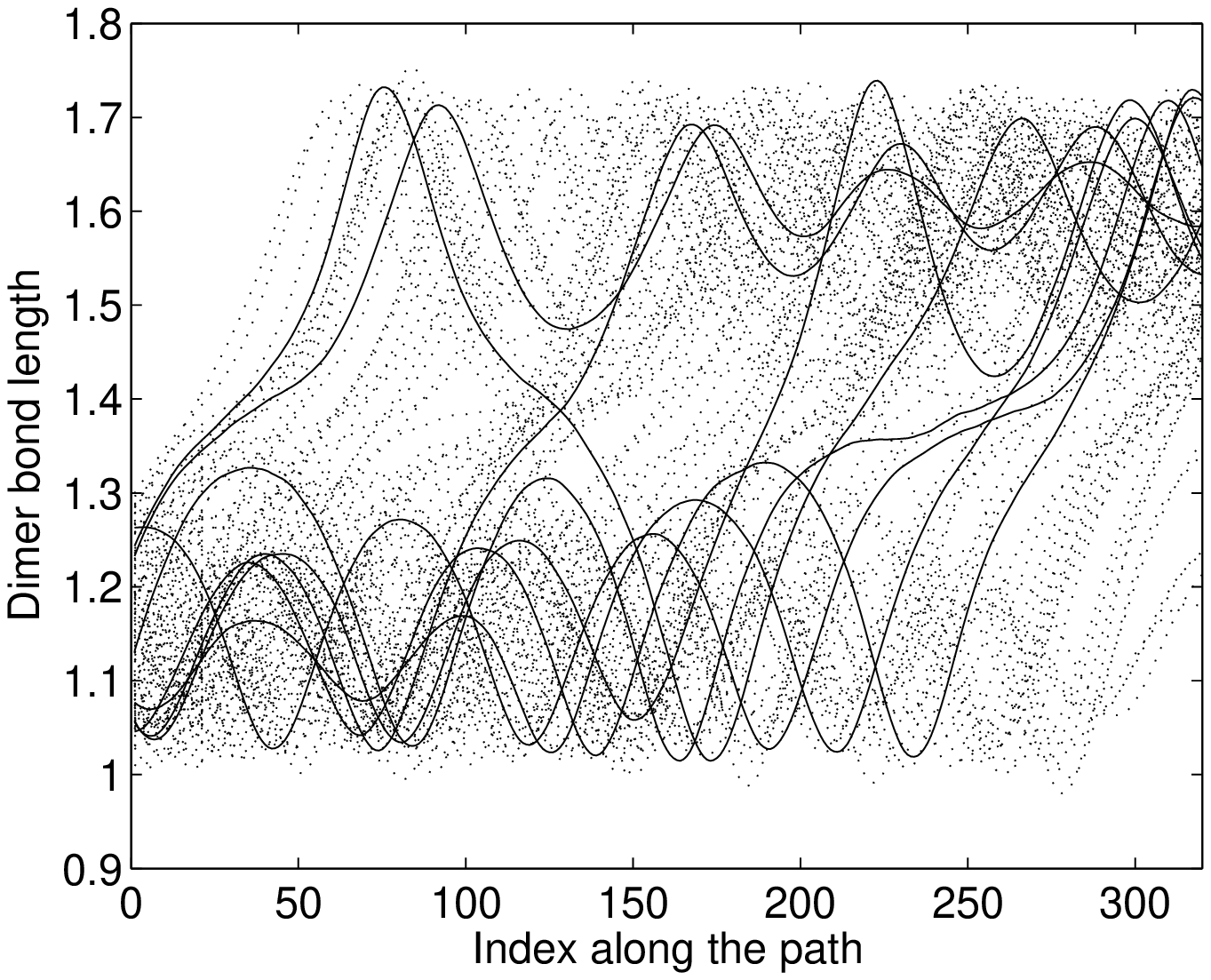}
\includegraphics[width=7.5cm]{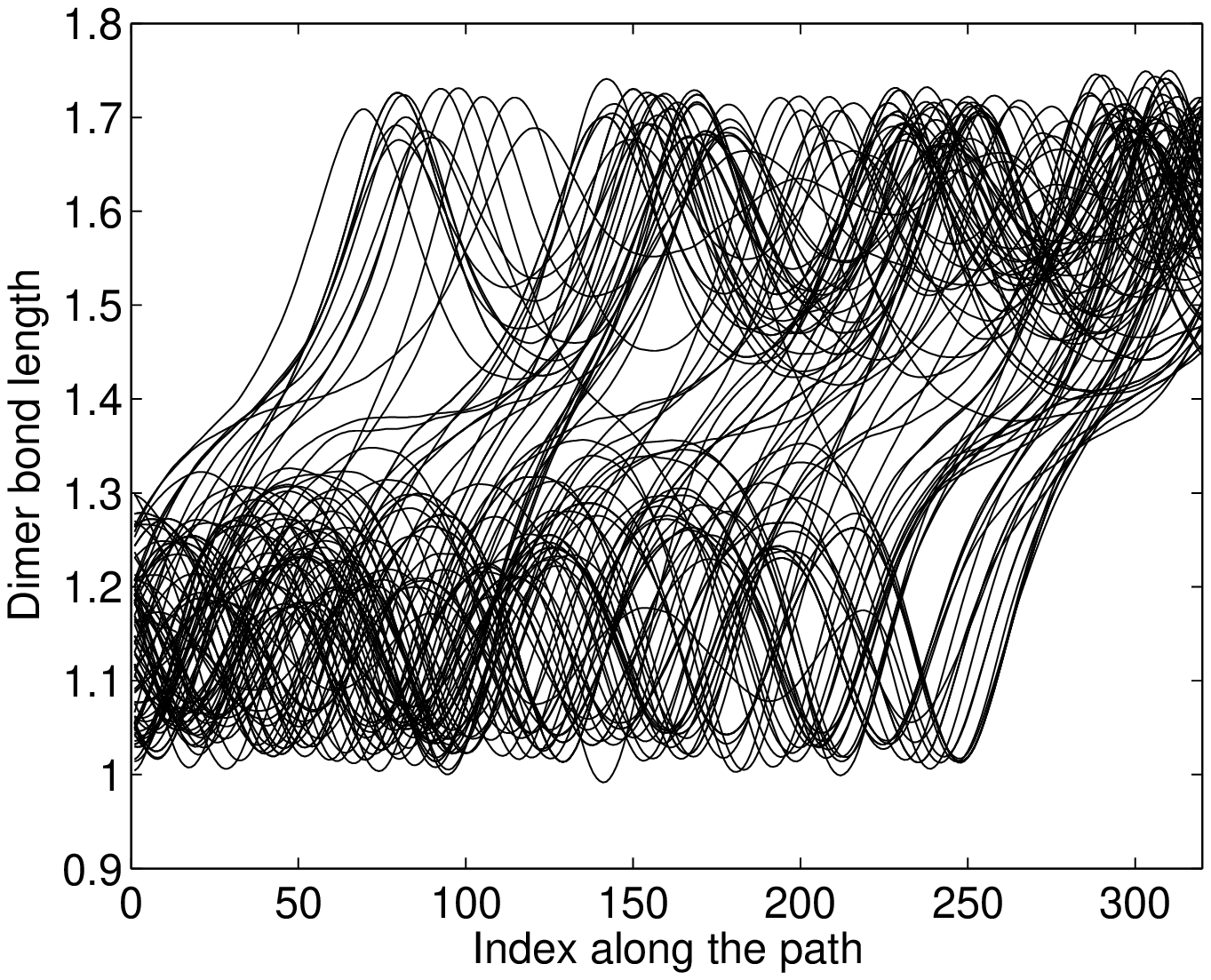}
\caption{\label{fig:sample} Comparison, for a nonequilibrium
switching of paths for $M=100$ systems in $n=1000$ steps without (Left) or with IPS (Right). Only
the paths having a weight greater than 0.05 are plotted in solid lines when plain nonequilibrium switching is used (the other paths are plotted in dotted lines).}
\end{figure}

We also present in Figure~\ref{fig:sample} a final sample computed using
a quite fast switching ($n=1000$) with a small sample of paths ($M=100$). Notice that all the 100 paths generated with the IPS switching are reactive, in contrast with the paths generated by a straightforward switching in the Jarzynski way. Besides, as a consequence of the degeneracy of paths, only 8 paths in 100 have a significant weight (larger than 0.05 when normalized by the total weight, as given by~(\ref{eq:weights})). This simple example shows why it is difficult to compute averages over the final sample of paths when performing plain nonequilibrium switching, and why it may be interesting to resort to some selection process to prevent such a degeneracy. 

In agreement with a previous study~\cite{RS06}, the results show that the IPS algorithm 
allows to reduce the variance on the estimates and to end up the simulation with a well-distributed and non-degenerate sample, provided the switching is slow enough.

\section{Conclusion and prospects}

In conclusion, we have presented here some new algorithms for path
sampling with stochastic dynamics, either equilibrium sampling (wich can
be used for the computation of free energy differences, or reaction
rates), or nonequilibrium sampling (which allows to perform simulated
annealing in a rigorous manner instead of performing simulated
tempering; or to switch from a sample of unconstrained paths to a sample
of constrained paths, and compute the associated ratio of partition functions).

The brownian tube proposal used for equilibrium sampling is a simple generalization of the previous approaches, and can therefore always be used as a shooting algorithm with only minor modifications to existing TPS algorithms. A systematic criterion for setting the correlation factors $\{\alpha_i\}_i$ would be to consider simple analytical forms as proposed at the end of section~\ref{sec:tube}, and choose $\{\alpha_i\}_i$ to obtain balanced acceptance/rejection rates or, when some specific observable has to be computed, to optimize the parameters to obtain the best convergence results (on some preliminary computations). However, for simulations of large systems using long paths, the brownian tube approach may be impossible to use because of the limited numerical precision and the chaotic behavior of the system: indeed, starting from a given path, it is not clear whether this path can be recovered by first computing the random numbers associated with the trajectory, and then integrating this trajectory again starting from the initial point.

The equilibration of the nonequilibrium switching dynamics is very intesting to reduce the variance of free energy computations when switching from unconstrained to constrained paths, or to obtain well-distributed ensemble of paths in the end (which is of paramount importance for the correctness of a simulated annealing procedure for example). However, the switching still has to be done slowly enough and using a number of replicas large enough. Once again, this may be problematic for very large systems.

It would be interesting now to extend the switching procedure to TIS~\cite{vEMB03,vEB05}, 
where the length of paths is not constant, but which is naturally
sequential in the way computations are done in practice: indeed, the
flux through the next intermediate interface is computed using a sample
of paths crossing the previous interface (this is the major difference
with the forward flux techniques of~\cite{AFR06} where only points on
the previous interface are kept).

\section*{Acknowledgements}

I warmly thank Peter Bolhuis, for fruitful discussions and stimulating
ideas (as well as an invitation to the University of Amsterdam where part of this work was
done). I also benefited from interesting discussions with Mathias
Rousset, Tony Lelièvre and Eric Cancès. This work was supported by the
ACI "Simulation Moléculaire" of the French Ministry of Research, and
initiated while I was a long term visitor of the Program "Bridging Time
and Length Scales" at IPAM (UCLA).

%
%

\section*{Appendix A: Specific Monte-Carlo moves for switching from 
unconstrained to constrained path ensembles}
\label{section_specific}

When an interpolating function $h_\lambda$ appearing 
in~(\ref{canonical_path_discretized_lambda}) (or, equivalently, some order
parameter $\xi$) is known, it is possible to
increase the likeliness of the end point of the trajectory by performing
a move on the last configuration in the direction opposite to $\nabla
h_\lambda(q)$ while keeping the random numbers used for the transitions.
These moves should of course be employed with other MC moves, especially
MC moves relying on some trajectory generation, in order to relax the
shift toward higher values of $h_\lambda$ or $\xi$.

More precisely, using for example an overdamped Langevin dynamics to
update the end configuration, the associated Metropolis-Hastings
Monte-Carlo elementary step is, starting from a path $x$ for a parameter
$\lambda$ (in the Langevin dynamics setting):
\newline

\begin{algo}
Starting from a path $x=(x_0,\dots,x_L)$,
\begin{itemize}
\item Compute the sequence of $2d$-dimensional noises $(\bar{U}_i)_{0 \leq i
\leq L-1}$ associated with the backward (time-reversed) integration
from $x_L$ to $x_0$;
\item Compute a final configuration as $y_L = x_L + \delta_\lambda
  \nabla \xi(q) + (2 \delta_\lambda / \beta)^{1/2} \, G$ where $G$
  is a $dN$-dimensional random gaussian vector;
\item Integrate the path backward (time-reversed), starting from $y_L$,
using the noises $(\bar{U}_i)_{0 \leq i \leq L-1}$ to obtain a path
$y=(y_0,\dots,y_L)$. The probabilty ${\cal P}(x,y)$ to obtain $y$ starting
from $x$ is therefore the probability to obtain $y_L$ from $x_L$, so
that
\[
{\cal P}(x,y)= {\rm p}_{\rm switch}(x_L,y_L) = \left ( \frac{\beta}{4 \pi \delta_\lambda}
\right )^{d/2} \exp \left ( -\frac{\beta}{4 \delta_\lambda^2}
  |y_L-x_L - \delta_\lambda \nabla \xi(q)|^2 \right ).
\]
\item Accept the new path $y$ with probability
\[
r(x,y) = \min \left ( 1, \frac{\pi(y){\cal P}(y,x)}{\pi(x){\cal P}(x,y)}
\right )= 
\min \left ( 1, \frac{\hA(y_0) \rho(y_0)}{\hA(x_0) \rho(x_0)}
\frac{{\rm p}_{\rm switch}(y_L,x_L)}{{\rm p}_{\rm switch}(x_L,y_L)} \right ).
\]
\end{itemize}
\end{algo}

The magnitude $\delta_\lambda$ can be made to depend a priori on $\lambda$. It is then adjusted in pratice 
on the fly by first computing the
values of the gradient for the endpoint of each replica, in order to ensure that the displacement is small enough.

%
%

\end{document}